\documentclass[12pt]{article}
\usepackage[margin=1in, paperwidth=8.5in, paperheight=11in]{geometry}
\usepackage{amsmath}
\usepackage{graphicx}%
\usepackage{amsfonts}%
\usepackage{amssymb}
\usepackage{color}
\usepackage{float,subfigure}
\usepackage{setspace}
\usepackage{comment}
\usepackage{setspace}     
\usepackage{chngpage}  
\usepackage{epsfig}
\usepackage{subfigure}
\usepackage{hyperref}
\usepackage{multirow}
\usepackage{ifsym}
\usepackage{caption}
\usepackage{wrapfig}
\usepackage{appendix}
\usepackage{layout}
\usepackage{titlesec}
\usepackage{rotating}
\usepackage{array,arydshln} 
\usepackage{amssymb}
\usepackage{pifont}
\usepackage[flushleft]{threeparttable}
\usepackage{enumitem}
\usepackage{flafter}

\titlespacing\section{0pt}{12pt plus 4pt minus 2pt}{0pt plus 2pt minus 2pt}
\titlespacing\subsection{0pt}{12pt plus 4pt minus 2pt}{0pt plus 2pt minus 2pt}

\date{}
\newcommand{\bbeta}{ \mbox{\boldmath $\beta$} }

\newcommand{\ind}{\perp\!\!\!\!\perp}

\usepackage[natbib=true,
    style=authoryear,
    sorting=none]{biblatex}
\numberwithin{equation}{section}

\begin{document}

\thispagestyle{empty}
\setcounter{page}{0}

\bigskip
\begin{center}
{\large \textbf{Biomarker-Guided Adaptive Enrichment Design with Threshold Detection for Clinical Trials with Time-to-Event Outcome}}

\bigskip
\bigskip

\textbf{Kaiyuan Hua$^{1}$, Hwanhee Hong$^{1}$, Xiaofei Wang$^{1}$}\\
$^{1}$ Department of Biostatistics and Bioinformatics, Duke University School of Medicine, Durham, North Carolina, 27705, USA \\

\singlespacing
\end{center}

\small
\doublespacing
\textsc{Abstract:} 
Biomarker-guided designs are increasingly used to evaluate personalized treatments based on patients' biomarker status in Phase II and III clinical trials. With adaptive enrichment, these designs can improve the efficiency of evaluating the treatment effect in biomarker-positive patients by increasing their proportion in the randomized trial. While time-to-event outcomes are often used as the primary endpoint to measure treatment effects for a new therapy in severe diseases like cancer and cardiovascular diseases, there is limited research on biomarker-guided adaptive enrichment trials in this context. Such trials almost always adopt hazard ratio methods for statistical measurement of treatment effects. In contrast, restricted mean survival time (RMST) has gained popularity for analyzing time-to-event outcomes because it offers more straightforward interpretations of treatment effects and does not require the proportional hazard assumption. This paper proposes a two-stage biomarker-guided adaptive RMST design with threshold detection and patient enrichment. We develop sophisticated methods for identifying the optimal biomarker threshold, treatment effect estimators in the biomarker-positive subgroup, and approaches for type I error rate, power analysis, and sample size calculation. We present a numerical example of re-designing an oncology trial. An extensive simulation study is conducted to evaluate the performance of the proposed design.

\bigskip

\textsc{Key words: Biomarker adaptive design; Continuous biomarker; Patient enrichment; Restricted mean survival time; Threshold detection. }

\clearpage
\section{Introduction} \label{section1}
Due to advancements in targeted therapies and precision medicine over the last two decades, there is a growing trend in the use of biomarker-guided designs for assessing personalized treatments in Phase II and Phase III clinical trials, especially in the field of oncology~\citep{simon2014biomarker,antoniou2016biomarker,antoniou2017biomarker}. Predictive biomarkers are often used in biomarker-guided trials to develop classifiers to identify appropriate patients as either excellent or poor candidates for clinical decisions to optimize therapy selections~\citep{lin2015reinventing,landeck2016biomarkers}. Recently, designs with adaptive enrichment have become increasingly attractive for biomarker-guided therapies as they provide additional flexibility during the trial~\citep{simon2013adaptive,antoniou2016biomarker}. In this paper, we consider the setting in which a single continuous biomarker is available at baseline. In general, a fixed threshold of the biomarker is pre-specified to dichotomize patients into ``biomarker-positive" and ``biomarker-negative" subgroups, and it is assumed that the biomarker-positive patients will benefit more from the new therapy over the active control. The biomarker-guided adaptive enrichment designs are typically conducted in a two-stage manner~\citep{frieri2023design,stallard2023adaptive}. In the first stage, they start with randomizing patients from the whole population. At the end of this stage, the accumulating data are used to decide whether to restrict accrual in the biomarker-positive subgroup in the second stage. Such adaptive enrichment could avoid further including patients who do not benefit from an intervention and may experience some negative effects~\citep{wang2018enrichment,lai2019adaptive,thall2021adaptive}. Existing literature demonstrates that such designs improve the efficiency of clinical trials by reducing the number of patients needed for randomization, particularly when the therapy of interest is known to be effective only in biomarker-positive patients and the biomarker assessment is highly accurate~\citep{wang2019enrich}.

However, the optimal cutoff for classifying the patients into positive and negative subgroups is often not available and challenging to be pre-specified at the design stage of a trial~\citep{simon2013adaptive}. To address this challenge, more flexible adaptive enrichment designs have been proposed by researchers~\citep{simon2013adaptive, simon2018using, wang2020design}. While the trial is ongoing, the observed data is used to identify the subgroup of patients most likely to benefit from a treatment (i.e., biomarker-positive subgroup) based on the biomarker values and certain utility functions. These designs allow simultaneously identifying the optimal biomarker threshold, defining the biomarker-positive subgroup, and estimating and testing the treatment effect in this subgroup. These approaches provide much efficiency of the enrichment strategy without the need to select a subset in advance, and the meaningful treatment effect estimates without being diluted by biomarker-negative patients who receive no or fewer benefits~\citep{wason2023discussion}.

Moreover, limited biomarker-adaptive enrichment designs have been proposed for time-to-event outcomes~\citep{mehta2014biomarker,diao2018biomarker,park2021bayesian}. These designs were developed based on hazard ratio (HR), which is commonly estimated by the Cox PH model~\citep{cox1972regression} or requires testing treatment effect with a log-rank test~\citep{bland2004logrank}. A fundamental assumption for Cox regression models is the PH assumption, which assumes that the HR between two groups does not vary over time. However, non-constant HRs have been recently observed in many oncology trials due to delayed treatment effects or other biomedical reasons~\citep{reck2016pembrolizumab,rittmeyer2017atezolizumab,barlesi2018avelumab}. In such scenarios, the PH assumption becomes problematic, rendering HR a potentially misleading and inappropriate summary of treatment effects~\citep{lin1989robust}. Violations of the PH assumption may also considerably impact the statistical power of the log-rank test to detect the treatment difference~\citep{mukhopadhyay2022log}. To address this issue, various alternative tests have been proposed~\citep{lin2020alternative}, and we emphasize a compelling alternative method based on Restricted Mean Survival Time (RMST)~\citep{uno2014moving,zhao2016restricted,tian2018efficiency} in this paper.

RMST summarizes the survival time up to a clinically relevant and usually pre-specified truncation time $t^*$. It is defined as the mean of the truncated event time $Y = \text{min}(T, t^*)$, and can be estimated by the area under the survival curve $S(t) = P[T>t]$ from $t = 0$ to $t = t^*$~\citep{royston2013restricted}:
\begin{equation} \label{eqn1.1}
\mu(t^*) = E(Y) = E[T \wedge t^*] = \int_0^{t^*}S(t)dt.
\end{equation}
The difference or ratio of RMST between two treatment groups measures the relative treatment effect concerning a gain or loss of event-free survival time up to $t^*$~\citep{kim2017restricted}. Estimating RMSTs does not necessitate model assumptions, and their interpretations are more straightforward than HRs across any distribution of time-to-event outcomes~\citep{perego2020utility}. Leveraging these advantages, many RMST-based clinical trial design methods have recently been proposed~\citep{trinquart2016comparison,weir2018design,luo2019design,lu2021statistical}. However, limited attention has been given to the biomarker-adaptive enrichment design. We address this methodological and practical gap by proposing a two-stage biomarker-guided adaptive RMST design with threshold detection and patient enrichment.

Our proposed design is motivated by a biomarker-guided randomized Phase III oncology trial, JAVELIN Lung 200~\citep{barlesi2018avelumab}. This trial compared the efficacy and safety of avelumab, an anti-PD-L1 immune checkpoint inhibitor antibody, with docetaxel, a standard care for patients with non-small-cell lung cancer (NSCLC). PD-L1 expression on tumor cells is a well-studied biomarker in NSCLC~\citep{sankar2022role}, and previous studies have indicated a positive association between high PD-L1 expression levels and progression-free survival (PFS) and overall survival (OS) after treatment with PD-L1 inhibitors for NSCLC~\citep{shi2020biomarkers}. In JAVELIN Lung 200, avelumab was not found to have lower OS compared to docetaxel within ``PD-L1-positive" patients, defined as patients with PD-L1 expression $\geq 1\%$, but exploratory analyses revealed that avelumab was associated with improved OS and PFS compared to docetaxel in patients with higher PD-L1 expression at $\geq 50\%$ and $\geq 80\%$ cutoffs. However, these exploratory analyses were post hoc, and the biomarker thresholds were pre-specified and thus possibly not optimal. Moreover, the PH assumption was violated for both OS and PFS (Supplemental Figures 1 and 3 in~\cite{barlesi2018avelumab}), so the characterization of the treatment effect based on the HR and its confidence interval estimated from the Cox PH model was not valid, and the hypothesis testing based on the log-rank test was less powerful. Consequently, there is a need for an RMST-based design to identify an optimal threshold and define the biomarker-positive subgroup while the trial is ongoing. Enrichment is enabled to increase the accrual of patients from the selected subgroup and enhance the design's efficiency and power.

\section{Design} \label{section2}
\subsection{Notations and Assumptions} \label{section2.1}
We consider a two-stage enrichment design where patients are recruited and randomized between experimental and control treatments. Let $Z$ be the treatment indicator, where $Z=1$ for the experimental and $Z=0$ for the control. Let $X \in [0,1]$ represent a single continuous biomarker of interest and is assumed to follow a standard uniform distribution, i.e., $X \sim \text{Unif}[0,1]$. If $X$ is a continuous marker with other distributions, the proposed design is still applicable when the values of $X$ can be scaled to $[0,1]$ through percentile transformations.

Following the potential outcomes framework~\citep{rubin1974estimating}, let $T(1)$ and $T(0)$ be the potential time-to-event under treatment and control, respectively. By assuming the consistency of potential outcomes, the time-to-event is $T = T(1)Z + T(0)(1-Z)$. Let $C$ be the censoring time, and we assume conditionally independent censoring given $Z$ and $X$, written as $C \ind T | Z, X$. In the presence of right censoring, we observe $U = \text{min}(T,C)$ and the censoring indicator $\delta = I[T \leq C]$, where $I[\cdot]$ is the indicator function.

We assume piecewise exponential hazard, $h_z(t|X)$, for patients with $Z=z$:
\begin{eqnarray} \label{eqn2.1}
    h_z(t|X) &=& \lambda_z(t) \times \text{exp}\{\gamma_zX\}, \mbox{ for } z = 0,1
\end{eqnarray}
where $\lambda_z(t)$ is the baseline hazard and $\gamma_z$ is the log-HR representing a constant association between biomarker and hazard. Consider $J_z$ time intervals for each treatment with change points $0 = \tau_{z,0} < \tau_{z,1} < \cdots < \tau_{z,J_z} = \infty$. We assume $\lambda_z(t)$ is constant within each interval, so that $\lambda_z(t) = \lambda_{z,j}, \mbox{ for } t \in [\tau_{z,j-1}, \tau_{z,j})$. Note that our proposed design is versatile and applicable to various hazard functions. The hazard function in Equation~\eqref{eqn2.1} is employed here for illustrative purposes, and alternative hazard models can be considered based on the requirement of a specific design.

\subsection{Optimal Biomarker Threshold} \label{section2.2}
Several methods have been proposed for selecting the biomarker threshold and identifying biomarker-positive subgroups~\citep{renfro2014adaptive,zhao2020designing,wang2020design}. For example, the subgroup can be selected by either maximizing the product of its sample size and effect size~\citep{zhao2020designing} or maximizing the test statistic of the treatment effect difference between two subgroups~\citep{renfro2014adaptive}. However, defining the best subgroup using these methods involves a trade-off between subgroup and treatment effect size~\citep{wang2020design}. For instance, selecting a smaller subgroup with a higher average treatment effect may not be the best choice. Furthermore, these approaches typically necessitate pre-specifying a set of candidate thresholds or employing a greedy search to identify the optimal one.

To address these issues, we adopt the methods introduced by~\cite{wang2020design} to identify the optimal threshold by the intersection point of the biomarker response curves of two treatment groups (i.e., the RMST curves as functions of biomarker value). This method parametrically models the biomarker response curves without mandating the pre-specification or greedy search for thresholds. Another advantage of using the intersection point is that it ensures the benefit of the experimental treatment for the biomarker-positive subgroup including all patients with positive treatment effects.

Here, the biomarker-positive subgroup comprises patients whose biomarker values are within the range where the conditional RMST differences between the experimental and control are all positive, given $X$. Conversely, the biomarker-negative subgroup comprises the range of $X$ where the conditional RMST differences are all negative. We refer to the threshold between biomarker-positive and negative subgroups as the biomarker cutpoint, denoted as $c_{t^*}$ concerning the truncation time $t^*$. Without loss of generality, we assume $X$ is positively associated with the RMST difference between treatments. As such, the biomarker-positive and negative subgroups are defined as $X \in (c_{t^*},1]$ and $X \in [0,c_{t^*})$, respectively.

We denote $\mu_z(t^*|X)$ as the conditional RMST function for $z \in \{0,1\}$, which is written as:
\begin{equation*}
    \mu_z(t^*|X) = \int_0^{t^*} S_z(t|X) dt = \int_0^{t^*} \text{exp} \left \{-\int_0^t h_z(u|X)du \right \} dt,
\end{equation*}
where $S_z(t|X)$ is the conditional survival function. The biomarker cutpoint, $X = c_{t^*}$, can be identified by solving the following equation concerning $X$:
\begin{equation} \label{eqn2.2}
\mu_1(t^*|X) - \mu_0(t^*|X) = 0.
\end{equation}
Note that $c_{t^*}$ may fall outside the range of $[0,1]$. In such instances, the biomarker cutpoint will be truncated at 0 or 1, which is formally written as $c_{t^*} = \text{min}\{\text{max}\{c_{t^*},0\},1\}$.

\subsection{Two-Stage Adaptive Enrichment Design} \label{section2.3}
\begin{figure}[!htbp]
\centerline{\includegraphics[width=6in]{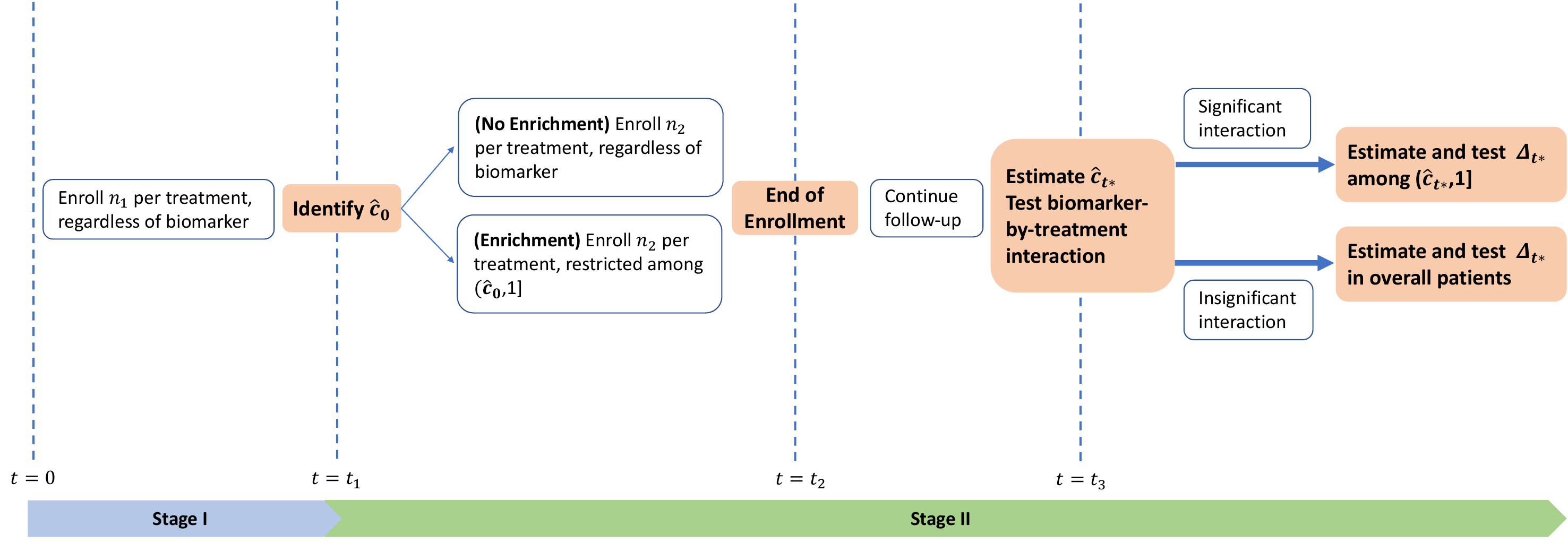}}
\caption[Design Diagram.]{Diagram of the two-stage adaptive and enrichment design.}
\label{fig1}
\end{figure}

Figure~\ref{fig1} visualizes the diagram of the proposed design. Stage I is conducted from the calendar time 0 to $t_1$, and accrues $n_1$ patients per treatment arm regardless of the biomarker value. At the end of Stage I (at $t_1$), a biomarker cutpoint, denoted as $\hat{c}_0$, is estimated concerning the RMST difference at a pre-specified truncation time $t^*$ related to a specific research question. Based on the result, two enrichment criteria can be introduced for Stage II: 1) accrue patients regardless of the biomarker value (i.e., no enrichment) in Stage II if $\hat{c}_0$ is not identified (e.g., all biomarker values are positive or negative to patients); 2) enrich and restrict the accrual in Stage II to patients in the biomarker-positive subgroup $(\hat{c}_0,1]$. Note that the trial will not be terminated in the first condition as the data are not mature in the early stage, and the treatment effect is not tested.

In Stage II, $n_2$ patients per treatment arm are accrued, and the enrollment continues from $t_1$ to $t_2$, resulting in a total of $n = 2n_1 + 2n_2$. All patients are continuously followed up until $t_3$, and we assume $t_3 > t^*$. After completing the follow-up, we estimate a biomarker cutpoint, denoted as $\hat{c}_{t^*}$, and assume a total of $n_+$ patients are included from both stages in the estimated biomarker-positive subgroup $(\hat{c}_{t^*},1]$. Additionally, we test whether there exists a statistically significant positive relationship between the biomarker and treatment effect. This hypothesis test, denoted as \textit{\textbf{Hypothesis Zero}}, can be written as follows:
\begin{itemize}
    \item Null hypothesis $H_{00}$: There is no positive biomarker-by-treatment interaction.
    \item Alternative hypothesis $H_{a0}$: There is a positive biomarker-by-treatment interaction.
\end{itemize}
Based on the testing result, two mutually exclusive hypotheses for conditional treatment effects (i.e., RMST difference) are considered. First, if $H_{00}$ is rejected, we estimate and test the RMST difference within $(\hat{c}_{t^*}, 1]$. This hypothesis test, denoted as \textit{\textbf{Hypothesis One}}, is written as:
\begin{itemize}
    \item Null hypothesis $H_{01}$: The RMST difference in the biomarker-positive subgroup $= 0$.
    \item Alternative hypothesis $H_{a1}$: The RMST difference in the biomarker-positive subgroup $> 0$.
\end{itemize}
Second, if $H_{00}$ is not rejected, we estimate and test the RMST difference in the overall population. This hypothesis test, denoted as \textit{\textbf{Hypothesis Two}}, is written as:
\begin{itemize}
    \item Null hypothesis $H_{02}$: The RMST difference in the overall population $= 0$.
    \item Alternative hypothesis $H_{a2}$: The RMST difference in the overall population $> 0$.
\end{itemize}
In Sections~\ref{section3} to~\ref{section5}, we will further describe the methods for identifying biomarker cutpoint, estimating treatment effects, and testing the aforementioned hypotheses.

\section{Biomarker Cutpoint Estimation} \label{section3}
We explore two methods for identifying the biomarker cutpoint: 1) the prediction method used in Stage I and 2) the RMST regression method used in Stage II.

\subsection{Prediction Method} \label{section3.1}
The prediction method involves two key steps. First, it fits a hazard function, $h_z(t|X)$, for each treatment based on the observed data. Second, it integrates the fitted hazard functions into Equation~\eqref{eqn2.2} and numerically estimates the biomarker cutpoint by solving this equation. This approach necessitates certain parametric assumptions on the survival curves or hazard functions within each treatment group. In this context, we focus on the piecewise exponential models, specified in Equation~\eqref{eqn2.1}.

For the $i^{th}$ patient and $j^{th}$ time interval, the hazard function can be written as follows:
\begin{eqnarray} \label{eqn3.1}
    \text{log}(\lambda_{z_i,ij}) = \text{log}(\lambda_{z_i,j}) + \gamma_{z_i}X_i, \mbox{ for } t \in [\tau_{z_i,j-1}, \tau_{z_i,j}),
\end{eqnarray}
where $z_i \in \{0,1\}$ is the treatment indicator and $\lambda_{z_i,j}$ is the baseline hazard in the $j^{th}$ interval. We use a Poisson log-linear regression method~\citep{holford1980analysis} to fit this model and estimate the parameters $\hat{\lambda}_{z,j}$ and $\hat{\gamma}_z$. The details are presented in Web Appendix A.

The strength of the prediction method lies in its ability to predict a biomarker cutpoint concerning a future truncation time $t^*$, which is particularly valuable in survival adaptive design where an early enrichment is sought to accrue more patients benefiting from the new therapy. In the context of our design, this approach is particularly advantageous in Stage I of our proposed design when the longest follow-up time $t_1$ is shorter than $t^*$. However, one limitation is its dependence on a parametric assumption on the survival curves. Moreover, under piecewise exponential models, it assumes no change point in the hazard functions between $t_1$ and $t^*$. As such, when a sufficient number of patients are being followed up beyond $t^*$, an alternative method (e.g., RMST regression method) can be used to estimate the biomarker cutpoint without the parametric model assumptions.

\subsection{RMST Regression Method} \label{section3.2}
The RMST regression method adapts the inverse probability of censoring weighted (IPCW) RMST regression~\citep{tian2014predicting} to estimate the biomarker cutpoint. We define the RMST up to $t^*$ given the biomarker and treatment as $\mu(t^{*}|X,Z) = E(Y|X,Z)$. The RMST regression model is written as follows:
\begin{eqnarray} \label{eqn3.2}
    g(\mu(t^{*}|X,Z)) = \beta_0 + \beta_1 Z + \beta_2 X + \beta_3 ZX,
\end{eqnarray}
where $g(\cdot)$ is the link function and identity link is used throughout this paper. The model parameters are denoted as $\hat{\bbeta} = [\hat{\beta}_0,\hat{\beta}_1,\hat{\beta}_2,\hat{\beta}_3]^T$. The conditional RMSTs given $X$ under the experimental and control treatments are $\hat{\mu}(t^{*}|X,1) = \hat{\beta}_0 + \hat{\beta}_1 + \hat{\beta}_2 X + \hat{\beta}_3 X$ and $\hat{\mu}(t^{*}|X,0) = \hat{\beta}_0 + \hat{\beta}_2 X$, respectively. Consequently, the conditional RMST difference is $\hat{\mu}(t^{*}|X,1) - \hat{\mu}(t^{*}|X,0) = \hat{\beta}_1 + \hat{\beta}_3 X$. Under the assumption that the treatment effect is positively related to the biomarker value, i.e., $\beta_3 > 0$, the biomarker cutpoint is $\hat{c}_{t^*} = -\hat{\beta}_1/\hat{\beta}_3$ with the biomarker-positive subgroup defined as $X \in [-\hat{\beta}_1/\hat{\beta}_3, 1]$.

The advantage of this approach is that it does not require parametric assumptions on the hazard functions. However, it is essential to ensure that the truncation time $t^*$ does not surpass the maximum of the observed event time (i.e., $U = \text{min}(T, C)$) within each treatment group. At the end of Stage II of the proposed design, we utilize this method to estimate the biomarker cutpoint as the final analysis time $t_3 > t^*$.

\section{Treatment Effect Estimators} \label{section4}
Let $c_{t^*}$ denote the true biomarker cutpoint, the treatment effect estimand, denoted as $\Delta_{t^*}^{(P)}$, is defined by the marginal RMST difference between the experimental and control treatments among the true biomarker-positive subgroup $(c_{t^*},1]$. It can be formally written as:
\begin{eqnarray} \label{eqn4.1}
   \Delta_{t^*}^{(P)} =  \int_{c_{t^*}}^1 \left \{ \mu_1(t^*|x) - \mu_0(t^*|x) \right \} dF_+(x),
\end{eqnarray}
where $\mu_1(t^*|x)$ and $\mu_0(t^*|x)$ are the conditional RMST for the experiment and control treatments, respectively, and $F_+(x)$ is the cumulative distribution function of $X$ within the true biomarker-positive subgroup. Note that $\Delta_{t^*}^{(P)} = \Delta_{t^*}^{(O)}$ when $c_{t^*}=0$ in Equation~\eqref{eqn4.1}.

\subsection{Naive Unadjusted Estimator} \label{section4.1}
The Naive unadjusted method employs the RMST definition in Equation~\eqref{eqn1.1} to estimate $\Delta_{t^*}$, written as follows:
\begin{eqnarray} \label{eqn4.2}
    \hat{\Delta}_1 = \int_0^{t^*} \left\{ \hat{S}_1(t) - \hat{S}_0(t)\right\} dt,
\end{eqnarray}
where $\hat{S}_z(t)$ is the standard Kaplan-Meier survival curve for Treatment $z$.  The estimator $\hat{\Delta}_1$ is consistent with $\Delta_{t^*}^{(P)}$ when the estimated biomarker cutpoint $\hat{c}_{t^*}$ is unbiased to $c_{t^*}$ and the biomarker distribution of $n_+$ patients aligns with $F_+(x)$. We discuss its large sample properties in Web Appendix B.1.

However, since the proposed design uses a different biomarker cutpoint $\hat{c}_0$ estimated at the end of Stage I for enrichment, the distribution of the included biomarker-positive patients may not align with the distribution for $X \in (\hat{c}_{t^*}, 1]$. Specifically, when $\hat{c}_0 \leq \hat{c}_{t^*}$, the biomarker value of the enriched patients in Stage II would cover the range of $X \in (\hat{c}_{t^*}, 1]$, while if $\hat{c}_0 > \hat{c}_{t^*}$, the biomarker values of the enriched patients represent only a subset of $(\hat{c}_{t^*}, 1]$. In such a situation, the distribution of those patients' biomarker would be truncated to $X \in (\hat{c}_0, 1]$, leading to an overestimation of $\hat{\Delta}_1$ under the positive assumption of biomarker-by-treatment interaction. As such, a weighting approach is needed to balance the distribution of the included $n_+$ biomarker-positive patients against the true distribution of $F_+(x)$.

\subsection{Calibration Weighted Estimators} \label{section4.2}
We use a calibration weighting (CW) method to balance the biomarker distribution. This method has been applied in causal inference for generalizing treatment effects from randomized trials to the target population~\citep{hainmueller2012entropy,josey2021transporting,lee2021improving}. It allows covariate distributions of the samples to empirically match a specific target population without fitting parametric models for the weight functions. In our proposed design, we leverage the CW method to generalize the RMST difference from the included positive patients to the true biomarker-positive subgroup with distribution $F_+(x)$. The calculation of the calibration weights $\hat{p}_i$ is introduced in Web Appendix B.2.

The CW-adjusted estimators of the RMST difference have been proposed by~\cite{hua2024inference}. Based on their approach, we propose four unbiased CW estimators of $\Delta_{t^*}$ in our two-stage adaptive enrichment design: CW Kaplan-Meier estimator ($\hat{\Delta}_2$), CW G-Formula Estimator ($\hat{\Delta}_3$), CW Hajek Estimator ($\hat{\Delta}_4$), and CW Augmented Estimator ($\hat{\Delta}_5$). The details and large sample properties of each estimator are provided in Web Appendix B.3.

\section{Type I Error Rate, Power, and Sample Size} \label{section5}
\subsection{Type I Error Rate and Critical Values} \label{section5.1}
We focus on the approaches that control the type I error rate under the global null setting such that $h_0(t|X)=h_1(t|X)$ for any $X \in [0,1]$. As outlined in Section~\ref{section2.3}, our proposed design incorporates one null hypothesis $H_{00}$ for testing positive biomarker-by-treatment interaction and two null hypotheses $H_{01}$ and $H_{02}$ for testing conditional treatment effects. We discuss the type I error rate controls on these two aspects.

First, we assign $\alpha_0$ as the significant level for testing $H_{00}$. Based on the RMST regression (Equation~\eqref{eqn3.2}) fitted at the end of Stage II, we evaluate significant positive biomarker-by-treatment interaction by testing if $\beta_3 > 0$. As such, the \textbf{\textit{Hypothesis Zero}} can be rewritten as: $H_{00}$: $\beta_3 = 0$ vs. $H_{a0}$: $\beta_3 > 0$. The test statistic is defined as $Z_{\beta_3} = \sqrt{n}\hat{\beta}_3/\hat{\sigma}_{\beta_3}$, where $\hat{\sigma}_{\beta_3}/\sqrt{n}$ is the standard error of $\hat{\beta}_3$. Under $H_{00}$, $Z_{\beta_3} \sim N(0,1)$. To control the type I error rate at $\alpha_0$, the critical value for rejecting $H_{00}$ is $q_0 = \Phi^{-1}(1-\alpha_0)$, where $\Phi^{-1}(\cdot)$ is the inverse cumulative distribution function of standard normal distribution.

Second, we define the family-wise type I error rate $\alpha$ for testing the treatment effect as the probability of rejecting $H_{01}$ or $H_{02}$ under the global null. We allocate $\alpha_1$ and $\alpha_2$ to test $H_{01}$ and $H_{02}$, respectively. Since $H_{01}$ and $H_{02}$ are mutually exclusive, we have $\alpha = \alpha_1 + \alpha_2$. We denote $Z_l^{(P)}$ and $Z_l^{(O)}$ as the test statistics when evaluating the treatment effect in the biomarker-positive and overall patients, respectively. According to the five treatment effect estimators proposed in Section~\ref{section4}, the test statistics are defined as $Z_l^{(P)} = \sqrt{n_+}\hat{\Delta}_l^{(P)}/\hat{\sigma}_l^{(P)}$ and $Z_l^{(O)} = \sqrt{n}\hat{\Delta}_l^{(O)}/\hat{\sigma}_l^{(O)}$, for $l = 1,...,5$. As such, $\alpha_1$ and $\alpha_2$ are calculated as follows:
\begin{eqnarray}
    \alpha_1 &=& P(H_{01} \text{ is rejected } | H_{00} \text{ is rejected} ) \times P( H_{00} \text{ is rejected } | \text{ global null}) \nonumber \\
    &=& P(Z_l^{(P)} > q | H_{00} \text{ is rejected} ) \times P( Z_{\beta_3} > q_0 | \text{ global null}) \nonumber \\
    &=& \pi_1 \times \alpha_0, \label{eqn5.1} \\
    \alpha_2 &=& P(H_{02} \text{ is rejected } | H_{00} \text{ is not rejected}) \times P( H_{00} \text{ is not rejected}| \text{ global null}) \nonumber \\
    &=& P(Z_l^{(O)} > q | H_{00} \text{ is not rejected}) \times P( Z_{\beta_3} \leq q_0 | \text{ global null}) \nonumber \\
    &=& \pi_2 \times (1-\alpha_0). \label{eqn5.2}
\end{eqnarray}
Here, $q$ is the critical value for testing the conditional treatment effects in both biomarker-positive and overall patients. We assume a significance level of $\Tilde{\alpha}$ for testing the conditional treatment effects, such that $q = \Phi^{-1}(1-\Tilde{\alpha})$. We let $\pi_1$ be the conditional error rate for $H_{01}$ when $H_{00}$ is rejected and $\pi_2$ be the conditional error rate for $H_{02}$ when $H_{00}$ is not rejected. When $H_{00}$ is not rejected, $Z_l^{(O)} \sim N(0,1)$, thus we have $\pi_2 = \Tilde{\alpha}$. When $H_{00}$ is rejected, a biomarker is estimated from the prediction method as introduced in Section~\ref{section3.1}, and we calculate $\pi_1$ through the Monte Carlo method. Taking JAVELIN Lung 200 as an example, $\pi_1$ can be approximated by bootstrapping the data from patients with docetaxel (the standard care) for both treatment groups. As a result, the family-wise type I error rate can be controlled at $\alpha$ by adjusting the values of $q$ and $q_0$.

\subsection{Global Power} \label{section5.2}
To calculate the power of our design, we consider a specific alternative setting with hazard functions $h_1(t|X)$ and $h_0(t|X)$. We denote $H_{a1}$ as an alternative setting such that a positive biomarker-by-treatment interaction exists and the RMST difference in the positive group is greater than 0. Once the critical values $q$ and $q_0$ are determined in Equations~\eqref{eqn5.1} and~\eqref{eqn5.2}, the global power can be calculated as:
\begin{eqnarray} \label{eqn5.3}
    \text{Global Power} &=& P(Z_l^{(P)} > q | H_{00} \text{ is rejected}, H_{a1} ) \times P( Z_{\beta_3} > q_0 | H_{a1}) \\
    &+& P(Z_l^{(O)} > q | H_{00} \text{ is not rejected}, H_{a1}) \times P( Z_{\beta_3} \leq q_0 | H_{a1}). \nonumber
\end{eqnarray}

\subsection{Sample Size Calculation} \label{section5.3}
Given the global power and the critical values $q$ and $q_0$, we calculate the sample size using Equation~\eqref{eqn5.3} and the asymptotic distribution of the proposed estimators (see Web Appendix B). Let $n_+^*$ and $n^*$ represent the required sample size in the biomarker-positive subgroup and the overall patients, respectively. Note that $n_+^* = f(n^*, c_{t^*})$ is a function of $n^*$ and $c_{t^*}$ based on the enrichment strategy and accrual rate in the proposed design. For example, in the all-comer design with no enrichment, $f(n^*, c_{t^*}) = n^*(1-c_{t^*})$.

Under $H_{a1}$, we define the true RMST difference in the biomarker-positive subgroup and the overall patients as $\Delta^{(P)}_{t^*}$ and $\Delta^{(O)}_{t^*}$, respectively. Then we have:
\begin{eqnarray*}
    \frac{\hat{\Delta}_l^{(P)}-\Delta^{(P)}_{t^*}}{\hat{\sigma}_l^{(P)}/\sqrt{n_+^*}} \sim N(0,1) \text{ and } \frac{\hat{\Delta}_l^{(O)}-\Delta^{(O)}_{t^*}}{\hat{\sigma}_l^{(O)}/\sqrt{n^*}} \sim N(0,1), \text{ for } l=1,\dots,5.
\end{eqnarray*}
To test the positive biomarker-by-treatment interaction, we have $\frac{\hat{\beta}_3 - \beta_3}{\hat{\sigma}_{\beta_3}/\sqrt{n^*}} \sim N(0,1)$ under $H_{a1}$. Therefore, $P( Z_{\beta_3} > q_0 | H_{a1}) = 1- \Phi(q_0 - \frac{\beta_3}{\hat{\sigma}_{\beta_3}/\sqrt{n^*}})$. Let $\eta = P( Z_{\beta_3} > q_0 | H_{a1})$, the Equation~\eqref{eqn5.3} can be written as:
\begin{eqnarray} \label{eqn5.4}
    \text{Global Power} = \eta \left[ 1 - \Phi(q - \frac{\Delta^{(P)}_{t^*}}{\hat{\sigma}_l^{(P)}/\sqrt{f(n^*,c_{t^*})}})\right]
    + (1-\eta) \left[ 1 - \Phi(q - \frac{\Delta^{(O)}_{t^*}}{\hat{\sigma}_l^{(O)}/\sqrt{n^*})}\right]
\end{eqnarray}
By solving this equation, we can calculate the required sample size to achieve the specific global power in our proposed design. The values of $\Delta^{(P)}_{t^*}$ and $\Delta^{(O)}_{t^*}$ can be theoretically derived from the hazard functions $h_1(t|X)$ and $h_0(t|X)$. Additionally, both $\sigma_l^{(P)}$ and $\sigma_l^{(O)}$ can be estimated by Monte Carlo method as described in~\cite{lu2021statistical}, details are provided in Web Appendix C.

\section{A Numerical Example} \label{section6}
In this section, we illustrate how a biomarker-guided adaptive trial with threshold detection and enrichment can be designed using our proposed methodology by redesigning the JAVELIN Lung 200 study. In the new design, we choose PFS as the endpoint of interest and use the RMST difference to quantify the treatment effects, which is justified by notable PH violations observed in the effect of avelumab. Necessary design parameters and assumptions are made in the new design to mimic the results reported in~\cite{barlesi2018avelumab}. We denote $Z=1$ to represent avelumab and $Z=0$ for docetaxel, the randomization ratio is one-to-one. The hazard functions for each treatment are assumed as follows:
\begin{eqnarray} \label{eqn6.1}
    h_0(t|X) &=& 2.5\text{log}(2), \nonumber \\
    h_1(t|X) &=& 
    \begin{cases}
      6\text{log}(2)\times \text{exp}\{-0.8X\} & \text{for $t \leq 1/6$} \\
      2\text{log}(2)\times \text{exp}\{-0.8X\} & \text{for $t > 1/6$}.
    \end{cases}
\end{eqnarray}
Here, time is measured in years. We designate the PD-L1 expression on tumor cells as the biomarker $X$. Similar to in JAVELIN Lung 200, we focus on the patients with PD-L1 expression $\geq 1\%$ in this design. As such, we assume $X \sim \text{Unif}(0.01, 1)$. As docetaxel is not a biomarker-related therapy, the hazard function for $Z=0$ is supposed to follow an exponential model and is unrelated to the biomarker. The hazard function for avelumab is assumed to be piecewise exponential, and according to Supplemental Figure 3 in~\cite{barlesi2018avelumab}, we posit a single change point at 2 months (i.e., $t=1/6$).

We assume patients are uniformly enrolled from time 0 to $t_1=0.5$ year in Stage I. In Stage II, patients are uniformly enrolled from $t_1$ to $t_2=1$ year, with restriction in the biomarker-positive subgroup. We assume equivalent accrual rates across both stages. All patients will be followed up until $t_3=2.5$ years. We assume the lost to follow-up time $L_i$ follows an exponential distribution with a rate $-\text{log}(1-0.05)/2$.

Treatment effect is measured as the RMST difference between avelumab and docetaxel up to $t^*=1.5$ years. Based on Equations~\eqref{eqn2.2} and~\eqref{eqn6.1}, the optimal biomarker cutpoint is $c_{t^*}=29.6\%$. Consequently, the marginal RMST difference among the biomarker-positive subgroup $(c_{t^*},1]$ and overall patients are $\Delta_{t^*}^{(P)}=0.137$ and $\Delta_{t^*}^{(O)}=0.082$ years, respectively.

We use Equation~\eqref{eqn5.4} to calculate the global power across a range of sample sizes $n^*$ under each proposed estimator. Given the equivalent accrual time and rates in two stages, $n_+^* = \frac{n^*}{2}(1 + \frac{1-c_{t^*}}{0.99})$. We plan to control the family-wise type I error rate at 2.5\%, and we allocate $\alpha_0 = 2.5\%$. The critical values are determined as $q_0=\Phi^{-1}(0.975)$ and $q=\Phi^{-1}(0.977)$ using a Monte Carlo method. The details can be found in Web Appendix C.

\begin{figure}
  \centerline{\includegraphics[width=5in]{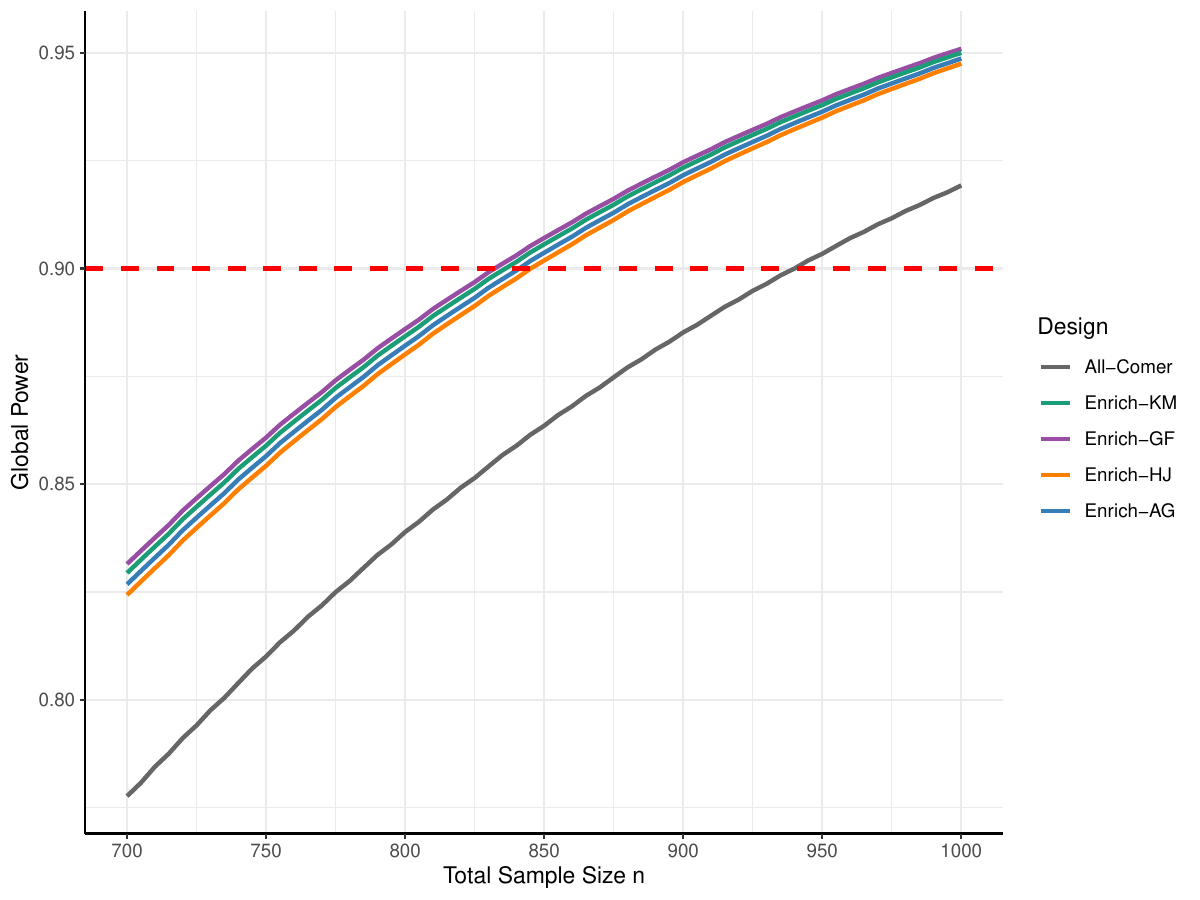}}
  \caption[Global power by total sample size in numerical example]{Curves of the global power by total sample size required in the all-comer design using naive unadjusted estimator and enrichment design using Kaplan-Meier (KM), G-Formula (GF), Hajek (HJ), and Augmented (AG) estimators.}
  \label{fig2}
\end{figure}

Figure~\ref{fig2} presents the curves of global power across a range of total sample sizes from the proposed enrichment design using the proposed estimators. Additionally, we compare the result from an all-comer design with no enrichment using the Naive unadjusted estimator. Notably, different estimators employed in the enrichment design yield very similar results. Compared to the all-comer design, the enrichment design requires fewer sample sizes to achieve the same global power. For example, to attain 90\% of the global power, the total sample size required in the all-comer design is $n=940$, while in the enrichment design with the Augmented estimator, it reduces to $n=845$. Furthermore, within the enrichment design, the G-Formula estimator demands the smallest sample size to achieve the same global power, while the Hajek estimator requires the largest sample size.

\section{Simulation Studies} \label{section7}
We conduct simulation studies to assess the properties of the proposed biomarker cutpoint estimating methods and treatment effect estimators within the framework of our proposed two-stage adaptive enrichment design. Additionally, we seek to evaluate the efficacy of the design and compare it with the all-comer design with no enrichment. Table~\ref{tab:1} describes four design scenarios in the simulation. Details of the data-generating mechanism, methods, and performance measures to assess operating characteristics are presented in Web Appendix D.

\begin{table}
\caption{Design scenarios in the simulation study. Enrichment designs are conducted using different settings in the prediction method to identify $\hat{c}_0$ at the end of Stage I. Different settings are regarding whether the number of change points in the piecewise exponential hazard functions and their locations are known or not.}
\label{tab:1}
\begin{center}
\begin{tabular}{cccc}
\hline
& & \multicolumn{2}{c}{Change Points}\\
Design Scenario & Description & Number & Locations \\
\hline
1 & All-comer design without & - & - \\
& enrichment & &\\
2 & Enrichment design using & Known & Known \\
& Prediction Method A & &\\
3 & Enrichment design using & Known & Unknown \\
& Prediction Method B & &\\
4 & Enrichment design using & Unknown & Unknown \\
& Prediction Method C & &\\
\hline
\end{tabular}
\end{center}
\end{table}

\subsection{Results for Biomarker Cutpoint Estimation}
Table~\ref{tab:2} summarizes the estimated biomarker cutpoints, $\hat{c}_0$ and $\hat{c}_{t^*}$, in Stages I and II under each design scenario. In Stage I, the Prediction Method A yields the most accurate prediction (bias = 0.001) since the number of change points and their locations in the piecewise exponential models are known. However, under Prediction Methods B and C with less known information on the change points, the bias of $\hat{c}_0$ increases. In Stage II, the RMST regression method yields unbiased estimates of $\hat{c}_{t^*}$ in all design scenarios regardless of the choice of the prediction method in Stage I.

\begin{table}
\caption{Estimated biomarker cutpoints, $\hat{c}_0$ and $\hat{c}_{t^*}$, in Stages I and II in the simulation study under four design scenarios. $\hat{c}_0$ is estimated by the prediction method and $\hat{c}_{t^*}$ is estimated by the RMST regression method. S.D.: standard deviation.}
\label{tab:2}
\begin{center}
\begin{tabular}{ccccccc}
\hline
& \multicolumn{3}{c}{Stage I} & \multicolumn{3}{c}{Stage II} \\
Design Scenario & $\hat{c}_0$ & Bias of $\hat{c}_0$ & S.D. of $\hat{c}_0$ & $\hat{c}_{t^*}$ & Bias of $\hat{c}_{t^*}$ & S.D. of $\hat{c}_{t^*}$ \\
\hline
1 & - & - & - & 0.522 & 0.003 & 0.062 \\
2 & 0.520 & 0.001 & 0.130 & 0.519 & 0.000 & 0.067 \\
3 & 0.506 & -0.013 & 0.132 & 0.520 & 0.001 & 0.066 \\
4 & 0.509 & -0.010 & 0.163 & 0.519 & 0.000 & 0.067 \\
\hline
\end{tabular}
\end{center}
\end{table}

\subsection{Results for Treatment Effect Estimation}
Table~\ref{tab:3} displays the estimated RMST differences in the biomarker-positive subgroup and their coverage probabilities under four design scenarios using the estimated biomarker cutpoint $\hat{c}_{t^*}$. In the all-comer design, all five estimators are unbiased with coverage probabilities over 98\%. In the enrichment designs, different prediction methods applied in Stage I do not alter the treatment effect estimations, as Design Scenarios 2-4 give similar results. Overall, the Naive unadjusted estimator overestimates the treatment effect, but all CW-adjusted estimators show unbiased results because the calibration weighting method balances the distribution of the included biomarker-positive patients to the true distribution of $F_+(x)$. The associated coverage probabilities are around 98\%. We also evaluate the estimated marginal RMST difference using the true biomarker cutpoint $c_{t^*}$, which helps verify the correctness of the proposed variance of each estimator. The results are presented in eTable 1 of the Supplementary Materials.

\begin{table}
\caption{Estimated marginal RMST differences in the estimated biomarker-positive subgroup under four simulation design scenarios, using the estimated biomarker cutpoint $\hat{c}_{t^*}$. RMSTD: marginal RMST difference in the estimated biomarker-positive subgroup, C.P.: coverage probability. $\hat{\Delta}_1$: Naive unadjusted estimator; $\hat{\Delta}_2$: CW Kaplan-Meier estimator; $\hat{\Delta}_3$: CW G-formula estimator; $\hat{\Delta}_4$: CW Hajek estimator; $\hat{\Delta}_5$: CW Augmented estimator.}
\label{tab:3}
\begin{center}
\begin{tabular}{ccccc}
\hline
Design Scenario & Estimator & Est. RMSTD & Bias of RMSTD & C.P. of RMSTD\\
\hline
1 & $\hat{\Delta}_1$ & 0.135 & 0.001 & 98.5\% \\
(All-Comer) & $\hat{\Delta}_2$ & 0.134 & 0.000 & 98.4\%\\
 & $\hat{\Delta}_3$ & 0.134 & 0.000 & 99.0\% \\
 & $\hat{\Delta}_4$ & 0.134 & 0.000 & 98.8\% \\
 & $\hat{\Delta}_5$ & 0.135 & 0.001 & 98.8\% \\
\hline
2 & $\hat{\Delta}_1$ & 0.144 & 0.010 & 96.5\% \\
(Enrichment) & $\hat{\Delta}_2$ & 0.134 & 0.000 & 97.8\%\\
 & $\hat{\Delta}_3$ & 0.135 & 0.001 & 98.5\% \\
 & $\hat{\Delta}_4$ & 0.135 & 0.001 & 98.2\% \\
 & $\hat{\Delta}_5$ & 0.135 & 0.001 & 98.1\% \\
\hline
3 & $\hat{\Delta}_1$ & 0.143 & 0.009 & 96.8\% \\
(Enrichment) & $\hat{\Delta}_2$ & 0.135 & 0.001 & 97.8\%\\
 & $\hat{\Delta}_3$ & 0.135 & 0.001 & 98.5\% \\
 & $\hat{\Delta}_4$ & 0.135 & 0.001 & 98.1\% \\
 & $\hat{\Delta}_5$ & 0.135 & 0.001 & 98.1\% \\
\hline
4 & $\hat{\Delta}_1$ & 0.145 & 0.010 & 95.4\% \\
(Enrichment) & $\hat{\Delta}_2$ & 0.135 & 0.001 & 97.5\%\\
 & $\hat{\Delta}_3$ & 0.135 & 0.001 & 98.2\% \\
 & $\hat{\Delta}_4$ & 0.135 & 0.001 & 97.8\% \\
 & $\hat{\Delta}_5$ & 0.135 & 0.001 & 97.8\% \\
\hline
\end{tabular}
\end{center}
\end{table}

\subsection{Results for Operating Characteristics}
Table~\ref{tab:4} summarizes the results from the power analysis. We first compare the average numbers of included negative patients in each design scenario. The all-comer design includes 1048 (51.9\%) true negative patients on average. However, the enrichment designs include significantly fewer negative patients. When less prior information on piecewise exponential hazard functions is used in the prediction method in Stage I, the average number of true negative patients increases from 628 (Design Scenario 2) to 653 (Design Scenario 4).

Subsequently, we compare the global power across five estimators. Within each design scenario, using the CW G-formula estimator, $\hat{\Delta}_3$, yields the largest global power due to its smallest variance, as outlined in eTable 1. Notably, the high global powers from the Naive unadjusted estimator, $\hat{\Delta}_1$, in the enrichment design stem from its overestimation of the RMST difference (see Table~\ref{tab:3}). The global powers from the other three CW-adjusted estimators are similar but lower than those from the CW G-formula estimator due to their larger variance. When considering the same estimator, the global power in the all-comer design is smaller than in the enrichment designs. Within the enrichment designs, the global powers are similar between using the first and second prediction methods (Design Scenario 2 and 3), and they are slightly higher than using the third prediction method (Design Scenario 4).

\begin{table}
\caption{Results from power analysis.}
\label{tab:4}
\begin{center}
\begin{tabular}{ccccccc}
\hline
Design & Avg. Number of & \multicolumn{5}{c}{Power of Using}\\
Scenario & True Negative Patients & $\hat{\Delta}_1$ & $\hat{\Delta}_2$ & $\hat{\Delta}_3$ & $\hat{\Delta}_4$ & $\hat{\Delta}_5$\\
\hline
1 & 1048 & 83.7\% & 80.0\% & 89.8\% & 77.4\% & 77.7\% \\
2 & 628 & 95.8\% & 90.2\% & 96.7\% & 88.9\% & 88.8\% \\
3 & 640 & 95.6\% & 90.6\% & 96.5\% & 89.1\% & 89.1\% \\
4 & 653 & 95.1\% & 88.5\% & 96.3\% & 86.8\% & 86.8\% \\
\hline
\end{tabular}
\end{center}
\end{table}

In eTables 2 and 3 of the Supplementary Materials, we present the type I error rates for testing \textbf{\textit{Hypothesis Zero}} and the family-wise type I error rate across various combinations of nominal significance levels of $\Tilde{\alpha}$ and $\alpha_0$ for each estimator under different design scenarios. Our findings demonstrate that the proposed design and estimators effectively retain a well-controlled family-wise type I error rate.

 \section{Discussion} \label{section8}
We presented a two-stage adaptive RMST design incorporating biomarker threshold detection and patient enrichment. Our approach involves two methods to identify the biomarker cutpoint and five estimators of the RMST difference in the biomarker-positive subgroup. Furthermore, we defined the family-wise type I error rate and global power in the context of our design and proposed a method for controlling the family-wise type I error rate under the global null by numerically determining the critical values for test statistics and a complementary method for calculating the sample size. We re-designed the JAVELIN Lung 200 using our design scheme and illustrated the sample size calculation process. Our findings demonstrated that the proposed enrichment design offers substantial reductions in the required sample size compared to the all-comer design while maintaining equivalent global power. Another important strength is its independence from needing a pre-specification of the biomarker-positive threshold, particularly for a continuous biomarker. Therefore, adopting the new design to oncology trials could avoid the potential failures of betting on an inaccurate biomarker cutpoint at the time of trial design, which is especially crucial when investigators lack sufficient data. Our flexible design allows adaptively identifying the threshold while simultaneously determining the biomarker-positive subgroup and estimating and testing the treatment effect in the ``selected" positive group. Furthermore, this RMST design applies to designs with time-to-event data when non-proportional hazards are expected.

Our theoretical work and extensive simulation study demonstrated that the proposed design and related methods addressed three key questions with satisfaction. First, in Stage I, the prediction methods can effectively identify the biomarker cutpoint for enrichment, and the accuracy increases as more prior information on the piecewise hazard functions becomes available. In Stage II, the RMST regression method provides an unbiased estimation of the biomarker cutpoint. Second, the calibration weighting method successfully balances the distribution of the included biomarker-positive patients with the true distribution of positive biomarkers. Consequently, the CW-adjusted estimators are all unbiased in the enrichment designs. The CW G-formula estimator demonstrates the smallest variance with the largest statistical power among all estimators. Third, compared to the all-comer design with an equivalent overall sample size, our proposed enrichment design features significantly more biomarker-positive patients on average and achieves higher global power while retaining a well-controlled family-wise type I error rate.

When outlining the proposed design for time-to-event outcomes, we primarily focused on utilizing the piecewise exponential model for hazard functions. This model is flexible as we allow an arbitrary number of change points, enabling us to approximate a wide range of event distributions. Notably, our proposed RMST regression method for estimating the biomarker cutpoint and the associated estimators for RMST difference does not necessitate the piecewise exponential hazard assumption. Additionally, it is crucial to underscore that our proposed design applies to other types of hazard functions, provided there is sufficient prior information on the parametric assumptions involved.

We focused on a practical scenario in the clinical trial with time-to-event endpoints, where the enrichment is implemented early to reduce the inclusion of biomarker-negative patients. The treatment effects are not tested at the end of Stage I as the follow-up time in this stage may not be sufficiently long. However, an additional statistical test can be introduced at the end of Stage I if the follow-up time exceeds the truncation time $t^*$ or if the RMST difference test uses a different truncation time earlier than $t^*$. Furthermore, the proposed design could be extended with a more complicated group-sequential RMST design~\citep{luo2019design,lu2021statistical}. This extension allows for incorporating $k \geq 1$ additional interim analysis in Stage II. At each interim analysis, enrichment can be simultaneously conducted with decisions on early termination for futility or efficacy. While our proposed design primarily considers a single truncation time, it is essential to note that challenges may arise when multiple truncation times for RMSTs are used to define treatment effects. For example, the multiplicity testing issues due to different treatment effect estimands, and the difficulties when identifying the biomarker cutpoint for enrichment under various truncation times. These topics remain for further research. Furthermore, future research can focus on evaluating the proposed design's performance regarding follow-up time and enrollment ratio between stages, exploring optimal or minimax designs, and extending from a single biomarker to multiple biomarkers for advanced methods in variable and cutpoint selection.

\section*{Acknowledgements}
Dr. Hong was partially supported by the National Institute of Mental Health (R01 MH126856) and the Patient-Centered Outcomes Research Institute (ME-2020C3-21145). Dr. Wang was partially supported by the NCI (P01 CA142538) and the NIA (R01 AG066883).

\section*{Code}
The relevant R code for the methodology, simulation study, and numerical example is available on is available on Github (\url{https://github.com/kimihua1995/RMST_DESIGN}).

\printbibliography

\end{document}


\begin{center}
{\large{ \textbf{Supplementary Material for ``Biomarker-Guided Adaptive Enrichment Design with Threshold Detection for Clinical Trials with Time-to-Event Outcome"}}} 
\end{center}

\bigskip

The supplementary material is organized as follows. In Web Appendix A, we outline the procedures for fitting the piecewise exponential hazard models utilized in the prediction method discussed in Section 3.1. In Web Appendix B, we show the calculation of the calibration weights $\hat{p}_i$ and introduce four CW-adjusted estimators within the framework of our proposed design. Web Appendix C presents the calculation of critical values $q$ and $q_0$ in the numerical example. In Web Appendix D, we describe the data-generating mechanism, methods, and performance measures to assess operating characteristics in the simulation study.

\section*{Appendix A: Modeling Fitting in Prediction Method}
For the $i^{th}$ patient and $j^{th}$ time interval, the piecewise exponential hazard function is written as follows:
\begin{eqnarray}
    \text{log}(\lambda_{z_i,ij}) = \text{log}(\lambda_{z_i,j}) + \gamma_{z_i}(1-x_i), \mbox{ for } t \in [\tau_{z_i,j-1}, \tau_{z_i,j}),
\end{eqnarray}
where $z_i \in \{0,1\}$ is the treatment indicator and $\lambda_{z_i,j}$ is the baseline hazard in the $j^{th}$ interval.

For patient $i \in \{1,\dots,n\}$, we first create some pseudo observations of $U_i$ and $\delta_i$ in each time interval, where $U_i = \text{min}(T_i, C_i)$ and $\delta_i = I[T_i \leq C_i]$. Let $u_{ij}$ denote the event free time by the $i^{th}$ patient in the $j^{th}$ interval, $[\tau_{z_i,j-1}, \tau_{z_i,j})$. If this patient has an event or is censored after $\tau_{z_i,j}$, then $u_{ij} = \tau_{z_i,j} - \tau_{z_i,j-1}$. Otherwise, if this patient has an event or is censored in this interval, $u_{ij} = U_i - \tau_{z_i,j-1}$. In addition, let $\delta_{ij}$ denote the censoring indicator in the $j^{th}$ time interval, such that $\delta_{ij} = 1$ if the $i^{th}$ patient has an event in this interval and $\delta_{ij} = 0$ otherwise. Then, the piecewise exponential model can be fitted to data by treating $\delta_{ij}$ as if they are independent Poisson observations with means $\mu_{ij} = u_{ij}\lambda_{z_i,ij}$. By replacing $\lambda_{z_i,ij} = \mu_{ij}/u_{ij}$, the Equation (S.1) can be written as follows:
\begin{eqnarray} 
    \text{log}(\mu_{ij}) = \text{log}(u_{ij}) + \alpha_{z_i,j} + \gamma_{z_i}(1-x_i),
\end{eqnarray}
where $\alpha_{z_i,j} = \text{log}(\lambda_{z_i,j})$. As such, the initial piecewise exponential model can be reformulated and adapted using the Poisson log-linear regression, which incorporates pseudo observations of $\{ \delta_{ij}, u_{ij}\}$ and employs $\text{log}(u_{ij})$ as an offset. Then the parameters $\hat{\lambda}_{z,j}$ and $\hat{\gamma}_z$ can be estimated from this Poisson regression model.

If the number of intervals $J_z$ and the change points $\tau_{z,0},\dots,\tau_{z,J_z}$ are known, fitting the model in Equation (S.2) is straightforward. However, without such prior information, we adopt a sequential testing approach~\cite{goodman2011detecting} to determine $J_z$ and estimate $\tau_{z,1},\dots,\tau_{z,J_z-1}$. Given a fixed number of intervals $J_z \geq 2$, we at first compute the maximum likelihood estimates for both the piecewise hazards $\alpha_{z,j}$ and change points $\tau_{z,j}$ for $j = 1,\dots,J_z-1$. Then, the process begins by testing the null hypothesis that there are no change points against the alternative with one change point. If the null hypothesis is rejected, we test the subsequent test involving the null hypothesis indicating one change point against the alternative with two change points in the model. This procedure continues until a null hypothesis cannot be rejected. They suggested using $\alpha^*(k) = \frac{\alpha^*}{2^{k-1}}$ for the $k^{th}$ test to control the overall type I error rate under $\alpha^*$~\cite{goodman2011detecting}.

It is essential to highlight that the assumption of piecewise exponential hazard function is necessary when employing the prediction method to identify the biomarker cutpoint. Furthermore, when adopting the approach outlined in~\cite{goodman2011detecting} to detect the number and location of the change points in the piecewise exponential hazard model, it is important to note that this approach assumes that the change points only affect the parameters of the baseline hazard function in Equation (2.1), that is, $\gamma_z$ remains constant over time. In situations where $\gamma_z$ varies over time, alternative methods can be employed, as proposed by~\cite{he2013sequential} and~\cite{zhang2014semiparametric}.

\section*{Appendix B: Calibration Weighting Method}
\subsection*{B.1: Naive Unadjusted Estimator}
Assume at the end of Stage II, there are a total of $n_+$ patients from two stages in the estimated biomarker-positive subgroup $(\hat{c}_{t^*}, 1]$. For $i = 1,...,n_+$ and $z = 0,1$, let $N_{zi}(t) = I[U_{i} \leq t; \delta_{i}=1; Z_{i}=z]$ denote the individual treatment-specific counting process and $Y_{zi}(t) = I[U_{i} \geq t; Z_{i}=z]$ denote the individual treatment-specific at-risk process. The treatment-specific KM estimator for the survival function is $\hat{S}_z(t) = \prod_{u \leq t} \left \{ 1 - \frac{dN_{z}(u)}{Y_{z}(u)}\right \}$, where $N_z(u) = \sum_{i=1}^{n_+} N_{zi}(u)$ and $Y_z(u) = \sum_{i=1}^{n_+} Y_{zi}(u)$. The Naive unadjusted RMST difference estimator is:
\begin{eqnarray} 
    \hat{\Delta}_1 = \int_0^{t^*} \left\{ \hat{S}_1(t) - \hat{S}_0(t)\right\} dt.
\end{eqnarray}
The estimator $\hat{\Delta}_1$ is consistent with $\Delta_{t^*}^{(P)}$ when the estimated biomarker cutpoint $\hat{c}_{t^*}$ is unbiased to $c_{t^*}$ and the biomarker distribution of $n_+$ patients aligns with $F_+(x)$. According to~\cite{zhao2016restricted,tian2018efficiency}, as the sample size $n_+ \to \infty$, $\sqrt{n_+}\{\hat{\Delta}_1 - \Delta_{t^*}^{(P)}\} \overset{d}{\to} N \left(0, \sigma_1^2 \right)$, where $\sigma_1^2$ is estimated by:
\begin{eqnarray*}
    \hat{\sigma}_1^2 = \int_0^{t^*} \left\{\int_t^{t^*}\hat{S}_1(u)du \right\}^2\frac{d\hat{\Lambda}_1(t)}{Y_1(t)/n_+} + \int_0^{t^*} \left\{\int_t^{t^*}\hat{S}_0(u)du \right\}^2\frac{d\hat{\Lambda}_0(t)}{Y_0(t)/n_+}.
\end{eqnarray*}
Here, $\hat{\Lambda}_z(t) = -\text{log}\{\hat{S}_z(t)\}$ is the Nelson-Aalen estimator for the cumulative hazard function for $z \in \{0,1\}$.

\subsection*{B.2: Calculation of Calibration Weights}
As patients in Stage I are accrued regardless of the biomarker value, we define the target population as the biomarker-positive subgroup in Stage I. Suppose that among the overall $n_+$ patients in the biomarker-positive subgroup, the first $\Tilde{n}_+$ patients are accrued from Stage I. For each patient $i=1,\dots,n_+$, the calibration weight $p_i$ is calculated through an optimization problem using the negative entropy objective function as follows:
\begin{equation*}
    min \left \{ \sum_{r=1}^{n_+}p_ilog(p_i) \right \},
\end{equation*}
with constraints:
\begin{eqnarray*}
\sum_{i=1}^{n_+}p_{i}\bg(x_i) &=& \Tilde{\bg} \\
\sum_{i=1}^{n_+}p_{i} &=& 1.
\end{eqnarray*}
Here, we let $\bg(x_i) = [x_i, x_i^2]^T$ and $\Tilde{\bg} = \sum_{i=1}^{\Tilde{n}_+}[x_{i}, x_{i}^2]^T/\Tilde{n}_+$. As such, the first constraint equalizes the first and second moments of $X$ between the overall biomarker-positive patients and the target population. It ensures that the weighted distribution of the biomarker among the positive patients in the proposed design empirically matches the distribution of $X$ from the target distribution. The second constraint implies that the calibration weights $p_i$ sum to a normalization constant of one, which guarantees a valid density function~\cite{hainmueller2012entropy,zhao2016entropy}.

The calibration weights $p_i$ can be calculated by using Lagrange Multiplier~\cite{de2000mathematical}. 
\begin{eqnarray}
    \hat{p}_i = \frac{exp\{\blambda^T\bg(\bX_i)\}}{\sum_{i=1}^{n_+}exp\{\blambda^T\bg(\bX_i)\}},
\end{eqnarray}
where $\blambda$ solves $\sum_{i=1}^{n_+}exp\{\blambda^T\bg(\bX_i)\}\{\bg(\bX_i) - \Tilde{\bg}\} = 0$.
\vspace*{10pt}

\subsection*{B.3: CW-Adjusted Estimators}
We propose four CW-adjusted estimators in the context of our proposed design. The structures of these estimators are similar to those proposed by~\cite{hua2024inference}. The proofs for the large sample properties of the proposed estimators can be found in the Supplementary Materials of~\cite{hua2024inference}.

To identify the estimand from the observed data in the proposed design, we make the following assumptions: 

\begin{description}
\item [Assumption 1 (Ignorability and positivity of trial assignment)] (i) $\{T(1),T(0)\} \indep Z | X $, and (ii) $0 < \pi(X) < 1$ with probability 1, where $\pi(X) = P(Z=1|X)$ is the treatment propensity score. In randomized trial, it can be commonly assumed to be constant, i.e., $\pi(X) = \pi$.

\item [Assumption 2 (Conditional noninformative censoring)] $\{T(1),T(0)\} \indep C | (Z, X) $, which implies $T \indep C | (Z, X)$.

\item [Assumption 3 (Covariate overlap with target population)] The true distribution of $X$ within the biomarker-positive subgroup, $F_+$, is absolutely continuous concerning the distribution of $X$ for the included biomarker-positive patients, denoted as $F^*$, i.e., for patients with biomarker value within $(\hat{c}_{t^*},1]$. That is, for any set of $X$, $A_{X}$, if $A_{X}$ has zero probability in $F^*$, then it also has zero probability in $F_+$. 

\item [Assumption 4 (Conditional exchangeability of survival function)] Assume that for any other uncalibrated covariates, denoted as $\bW$, $S(t|X,Z,\bW) = S(t|X,Z)$. This assumption also implies the conditional exchangeability of RMST.
\end{description}

According to Assumptions 1 to 4, $\Delta_{t^*}^{(P)}$ can be identified by $E[\{\mu_1(t^*|X) - \mu_0(t^*|X) \} \frac{dF_+}{dF^*}(X)]$, where $\frac{dF_+}{dF^*}(X)$ is the ratio of the probability density functions between the included biomarker-positive patients and the true distribution of the biomarker-positive subgroup.

\subsubsection*{Calibration Weighted Kaplan-Meier Estimator}
For $z \in \{0,1\}$, let $\Tilde{N}_z(t) = \sum_{i=1}^{n_+}\hat{p}_iN_{zi}(t)$ denote the CW treatment-specific counting process and $\Tilde{Y}_z(t) = \sum_{i=1}^{n_+}\hat{p}_iY_{zi}(t)$ denote the CW treatment-specific at-risk process, the CW Kaplan-Meier estimator is written as:
\begin{eqnarray} 
    \hat{\Delta}_2 = \Tilde{\mu}_1(t^*) - \Tilde{\mu}_0(t^*),
\end{eqnarray}
where $\Tilde{\mu}_z(t^*)$ is the treatment-specific CW RMST, defined as:
\begin{eqnarray*}
    \Tilde{\mu}_z(t^*) = \int_0^{t^*}\Tilde{S}_z(t)dt = \int_0^{t^*} \prod_{u \leq t} \left \{ 1 - \frac{d\Tilde{N}_z(u)}{\Tilde{Y}_z(u)}\right \} dt.
\end{eqnarray*}

According to Theorem 1 in~\cite{hua2024inference}, $\hat{\Delta}_2$ is consistent with $\Delta_{t^*}^{(P)}$, and as the sample size $n_+ \to \infty$, $\sqrt{n_+}\{\hat{\Delta}_2 - \Delta_{t^*}^{(P)}\} \overset{d}{\to} N \left(0, \sigma_2^2 \right)$, where $\sigma_2^2$ is estimated by:
\begin{eqnarray*}
    \hat{\sigma}_2^2 &=& n_+ \sum_{z=0,1} \int_0^{t^*} \left \{ \int_u^{t^{*}}\Tilde{S}_{z}(t)dt \right \}^2 \frac{d\Tilde{N}_{z}(u)}{\Tilde{W}_{z}(u)(\Tilde{Y}_{z}(u) - \Delta \Tilde{N}_{z}(u))},
\end{eqnarray*}
where $\Tilde{W}_{z}(u) = \Tilde{Y}_{z}^2(u)/\sum_{i=1}^{n_+}\hat{p}_i^2Y_{zi}(u)$ and $\Delta \Tilde{N}_{z}(u) = \Tilde{N}_{z}(u) - \Tilde{N}_{z}(u-)$.

\subsubsection*{Calibration Weighted G-Formula Estimator}
The CW G-Formula estimator employs the G-computation technique~\cite{robins1986new,robins2008estimation,naimi2017introduction}, which combines the IPCW RMST regression as introduced in Section 3.2 and the calibration weights. This estimator is a direct regression estimator and its outcome model is identified by the conditional RMSTs defined in Section 3.2. Here, we let the outcome models be $m_1(X) = \hat{\mu}(t^*|X,1)$ and $m_0(X) = \hat{\mu}(t^*|X,0)$. The CW G-formula estimator is written as:
\begin{eqnarray}
    \hat{\Delta}_3 = \frac{\sum_{i=1}^{n_+} \hat{p}_i\{m_1(x_i)-m_0(x_i)\}}{\sum_{i=1}^{n_+} \hat{p}_i} = \sum_{i=1}^{n_+} \hat{p}_i(\hat{\beta}_1 + \hat{\beta}_3 x_i)
\end{eqnarray}
According to Theorem 2 in~\cite{hua2024inference}, when the outcome models are not misspecified, $\hat{\Delta}_3$ is consistent with $\Delta_{t^*}^{(P)}$. As the sample size $n_+ \to \infty$, $\sqrt{n_+}\{\hat{\Delta}_3 - \Delta_{t^*}^{(P)}\} \overset{d}{\to} N \left(0, \sigma_3^2 \right)$, where $\sigma_3^2$ can be estimated using the Delta method~\cite{dowd2014computation}:
\begin{eqnarray*}
    \hat{\sigma}_3^2 = n_+ J_{\hat{\beta}}^T \Sigma_{\hat{\beta}}J_{\hat{\beta}},
\end{eqnarray*}
where $\Sigma_{\hat{\beta}} = \begin{pmatrix}
\mbox{Var}(\hat{\beta}_1) & \mbox{Cov}(\hat{\beta}_1, \hat{\beta}_3) \\
\mbox{Cov}(\hat{\beta}_1, \hat{\beta}_3) & \mbox{Var}(\hat{\beta}_3)
\end{pmatrix}$ is the variance-covariance matrix of the parameter vector $[\hat{\beta}_1, \hat{\beta}_3]^T$ and 
\begin{equation*}
    J_{\hat{\beta}} =  \sum_{i=1}^{n_+} \frac{\partial \hat{p}_i(\hat{\beta}_1 + \hat{\beta}_3 x_i)}{\partial [\hat{\beta}_1, \hat{\beta}_3]^T}  = [ 1, \sum_{i=1}^{n_+} \hat{p}_ix_i ]^T.
\end{equation*}

\subsubsection*{Calibration Weighted Hajek Estimator}
The CW Hajek estimator combines the calibration weights and another estimator of RMST, $\mu(t^*)$, proposed by~\cite{bang2000estimating} using the inverse probability of censoring weight (IPCW). For $i=1,\dots,n$, let $Y_i = \text{min}(T_i,t^*)$ denote the individual truncated time-to-event outcome at $t^*$. In the presence of right censoring, the IPCW estimator of $\mu(t^*)$ is
\begin{eqnarray*} 
    \hat{\mu}_{IPCW}(t^*) = \frac{1}{n}\sum_{i=1}^n\frac{\delta_i^*}{\hat{G}(Y_i)}Y_i,
\end{eqnarray*}
where $\delta_i^* = I[C_i \geq Y_i]$, and $\hat{G}(Y_i) = P(C_i > Y_i)$ is the Kaplan-Meier estimator of the survival function for the censoring time based on $\{(U_i, 1-\delta_i), i=1,\dots,n\}$. To understand the validity of $\hat{\mu}_{IPCW}(t^*)$, note that
\begin{eqnarray*}
    E \left[ \frac{\delta_i^*}{\hat{G}(Y_i)}Y_i | T_i \right] = Y_i\frac{P(C_i \geq Y_i|T_i)}{P(C_i > Y_i)} = Y_i.
\end{eqnarray*}
As such, the CW Hajek estimator is defined as:
\begin{eqnarray}
    \hat{\Delta}_4 = \frac{\sum_{i=1}^{n_+}\hat{p}_iZ_iw_iY_i}{\sum_{i=1}^{n_+}\hat{p}_iZ_iw_i} - \frac{\sum_{i=1}^{n_+}\hat{p}_i(1-Z_i)w_iY_i}{\sum_{i=1}^{n_+}\hat{p}_i(1-Z_i)w_i},
\end{eqnarray}
where $w_i = \delta_i^*/\hat{G}(Y_i)$ is the IPCW for the $i^{th}$ patient. According to Theorem 3 in~\cite{hua2024inference}, $\hat{\Delta}_4$ is consistent with $\Delta_{t^*}^{(P)}$ and as $n_+ \to \infty$, $\sqrt{n_+}\{\hat{\Delta}_4 - \Delta_{t^*}^{(P)}\} \overset{d}{\to} N \left(0, \sigma_4^2 \right)$.

The large sample properties of this estimator can be derived from the M-estimator theory~\cite{stefanski2002calculus}. Let $\btheta = [\theta_0, \theta_1]^T$ as the collection of parameters to be estimated. Then $\hat{\Delta}_4 = \hat{\theta}_1 - \hat{\theta}_0$ jointly solves the estimation equations as follows:
\begin{eqnarray*}
    \sum_{i=1}^{n_+}\bPhi_i^{H}(\btheta) = \sum_{i=1}^{n_+} \begin{pmatrix}
\hat{p}_iZ_iw_i(Y_i - \theta_1) \\
\hat{p}_i(1-Z_i)w_i(Y_i - \theta_0) 
\end{pmatrix} = 0.
\end{eqnarray*}
The sandwich variance estimator for $\hat{\btheta}$ is:
\begin{eqnarray*}
    \Sigma_{\hat{\btheta}} = \left\{ \sum_{i=1}^{n_+} \frac{\partial \bPhi_i^{H}(\hat{\btheta})}{\partial \btheta^T}\right\}^{-1} \left\{ \sum_{i=1}^{n_+} \bPhi_i^{H}(\hat{\btheta}) \bPhi_i^{H}(\hat{\btheta})^T \right\} \left\{ \sum_{i=1}^{n_+} \frac{\partial \bPhi_i^{H}(\hat{\btheta})}{\partial \btheta^T}\right\}^{-1}.
\end{eqnarray*}
By the M-estimator theory, $\hat{\btheta}$ is consistent to $\btheta$, and as the sample size $n_+ \to \infty$, $\sqrt{n_+}\{\hat{\btheta} - \btheta\} \overset{d}{\to} N \left(0, n_+\Sigma_{\hat{\btheta}} \right)$. Then by continuous mapping theorem,  $\sqrt{n_+}\{\hat{\Delta}_4 - \Delta_{t^*}^{(P)}\} \overset{d}{\to} N \left(0, \sigma_4^2 \right)$, where $\sigma_4^2$ can be estimated by $n_+[1,-1]\Sigma_{\hat{\btheta}}[1,-1]^T = n_+[\mbox{Var}(\hat{\theta}_1) + \mbox{Var}(\hat{\theta}_0) - 2\mbox{Cov}(\hat{\theta}_1,\hat{\theta}_0)]$.

\subsubsection*{Calibration Weighted Augmented Estimator}
The CW augmented estimator combines the CW Hajek and CW G-formula estimators:
\begin{eqnarray}
    \hat{\Delta}_5 &=& \frac{\sum_{i=1}^{n_+}\hat{p}_iZ_iw_i\{Y_i- m_1(x_i)\}}{\sum_{i=1}^{n_+}\hat{p}_iZ_iw_i} - \frac{\sum_{i=1}^{n_+}\hat{p}_i(1-Z_i)w_i\{Y_i - m_0(x_i)\}}{\sum_{i=1}^{n_+}\hat{p}_i(1-Z_i)w_i} \nonumber \\ 
    &+& \frac{\sum_{i=1}^{n_+}\hat{p}_i\{m_1(x_i) - m_0(x_i)\}}{\sum_{i=1}^{n_+}\hat{p}_i}.
\end{eqnarray}
Based on the semiparametric theory~\cite{tsiatis2006semiparametric}, this estimator is doubly robust and does not require the correct specification of the outcome models. According to Theorem 4 in~\cite{hua2024inference}, $\hat{\Delta}_5$ is consistent with $\Delta_{t^*}^{(P)}$. As the sample size $n_+ \to \infty$, $\sqrt{n_+}\{\hat{\Delta}_5 - \Delta_{t^*}^{(P)}\} \overset{d}{\to} N \left(0, \sigma_5^2 \right)$.

By the M-estimator theory, let $\bnu = [\nu_0, \nu_1, \nu_2]^T$ denote as the collection of parameters to be estimated. Then $\hat{\Delta}_5 = \hat{\nu}_1 - \hat{\nu}_0 + \hat{\nu}_2$ jointly solves the following estimation equations:
\begin{eqnarray*}
    \sum_{i=1}^{n_+}\bPhi_i^{A}(\bnu) = \sum_{i=1}^{n_+} \begin{pmatrix}
    \hat{p}_iZ_iw_i\{Y_i - m_1(x_i) - \nu_1\} \\
    \hat{p}_i(1-Z_i)w_i\{Y_i - m_0(x_i) - \nu_0\} \\
    \hat{p}_i\{m_1(x_i) - m_0(x_i) - \nu_2\}
\end{pmatrix} = 0.
\end{eqnarray*}
The sandwich variance estimator for $\hat{\bnu}$ is:
\begin{eqnarray*}
    \Sigma_{\hat{\bnu}} = \left\{ \sum_{i=1}^{n_+} \frac{\partial \bPhi_i^{A}(\hat{\bnu})}{\partial \bnu^T}\right\}^{-1} \left\{ \sum_{i=1}^{n_+} \bPhi_i^{A}(\hat{\bnu}) \bPhi_i^{A}(\hat{\bnu})^T \right\} \left\{ \sum_{i=1}^{n_+} \frac{\partial \bPhi_i^{A}(\hat{\bnu})}{\partial \bnu^T}\right\}^{-1}.
\end{eqnarray*}
As the sample size $n_+ \to \infty$, $\sqrt{n_+}\{\hat{\Delta}_5 - \Delta_{t^*}^{(P)}\} \overset{d}{\to} N \left(0, \sigma_5^2 \right)$, where $\sigma_5^2$ can be estimated by $n_+[1,-1,1]\Sigma_{\hat{\bnu}}[1,-1,1]^T$.

\section*{Appendix C: Supplement for Numerical Example}
eFigure 1 displays the survival curves derived from the simulated data among patients with PD-L1 expression $\geq 1\%$ and $\geq 50\%$. These curves closely resemble those depicted in~\cite{barlesi2018avelumab}, where the PH assumption is violated.

\begin{figure}[!htbp]
\centering\includegraphics[scale = 0.5]{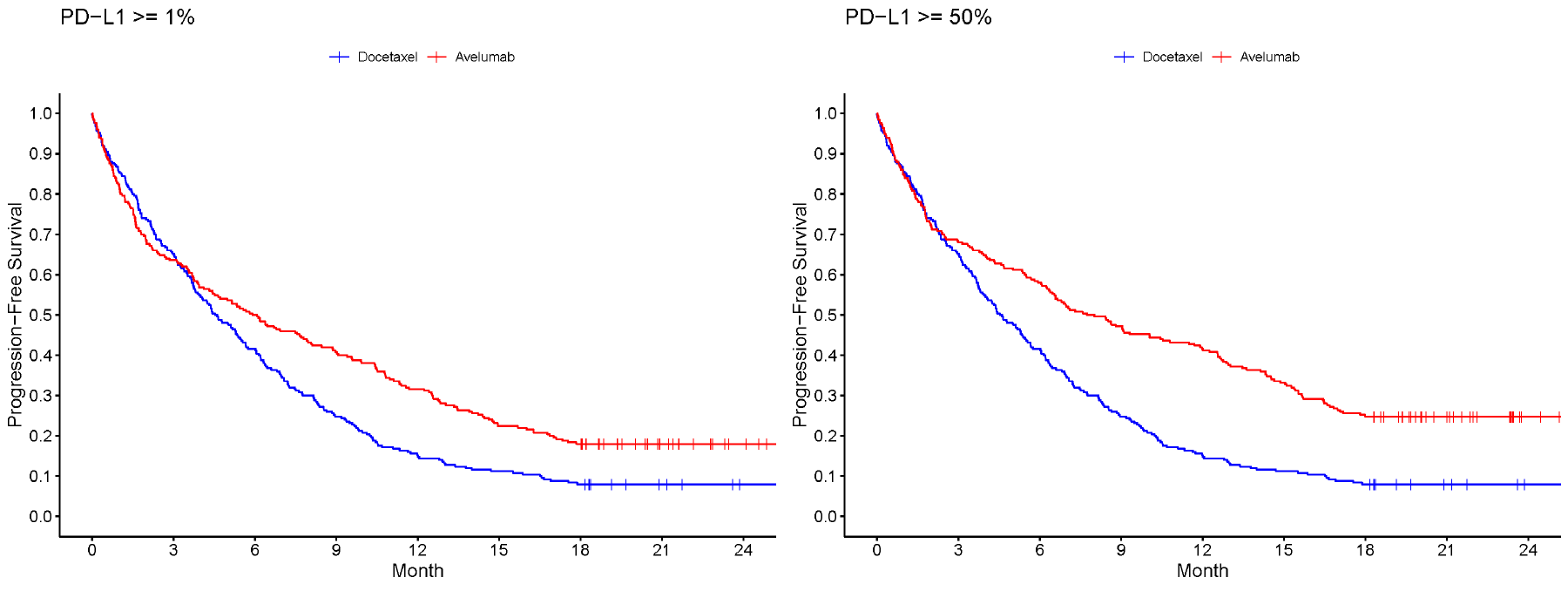}
\caption{Simulated RMST curves of progression-free survival in patients with PD-L1 $\geq 1\%$ (left panel) and PD-L1 $\geq 50\%$ (right panel).}
\end{figure}

Regarding the CW-adjusted estimators, it is essential to note that during the design for sample size calculation, we use the true biomarker cutpoint for enrichment and determining the biomarker-positive subgroup, i.e., $\hat{c}_0 = \hat{c}_{t^*} = c_{t^*}$. Consequently, we can assume the distribution of the included biomarker-positive patients aligns with the true distribution of $F_+(x)$. In such a situation, utilizing the calibration weighting method to balance distributions becomes unnecessary, and we can set $\hat{p}_i=1$ in each CW-adjusted estimator. By doing so, the CW Kaplan-Meier estimator degenerates to the Naive unadjusted estimator.

We plan to control the family-wise type I error rate at 2.5\%, and we allocate a significance level of 2.5\% for testing positive biomarker-by-treatment interaction (i.e., $\alpha_0=2.5\%$ for testing $H_{00}$). Next, we use the Monte Carlo method to determine the significance level of $\Tilde{\alpha}$ for testing the conditional treatment effects (i.e., $H_{10}$ and $H_{20}$). We conduct a simulation under the global null setting where $h_1(t|X) = h_0(t|X) = 2.5\text{log}(2)$ with a total sample size $n=10,000$ across a range of $\Tilde{\alpha} \in \{1.5\%, 1.6\%, ..., 2.5\%\}$. Based on the testing results from 10,000 simulated datasets, the family-wise type I error rate can be controlled at 2.5\% with $\Tilde{\alpha} = 2.3\%$. Therefore, the critical values used in Equation (5.4) are $q_0=\Phi^{-1}(0.975)$ and $q=\Phi^{-1}(0.977)$. Additionally, the asymptotic variance $\hat{\sigma}_{\beta_3}^2$, $[\hat{\sigma_l}^{(P)}]^2$, and $[\hat{\sigma_l}^{(O)}]^2$ in Equation (5.4) are estimated by Monte Carlo method as introduced in~\cite{lu2021statistical} with a total sample size $n=10,000$ and 10,000 replicates. Take $\sigma_l^{(O)}$ for example, one may first simulate a data set of a large sample size $M$, and then calculate the centered RMST estimates, $I = \sqrt{M}\left\{ \hat{\Delta}_l^{(O)} - \Delta^{(O)}_{t^*} \right\}$. After repeating this process a large number of $B$ times, $\hat{\sigma}_l^{(O)}$ can be approximated by $\sqrt{\sum_{b=1}^B I_b^2/B}$. As an illustration, this simulation utilizes the naive estimator (i.e., $l=1$), but other estimators can also be employed. However, the result might be slightly different due to the different variance of each estimator.

\section*{Appendix D: Simulation Studies}
\subsection*{D.1: Aims}
The simulation study aims to assess the properties of the proposed biomarker cutpoint estimating methods and treatment effect estimators within the framework of our proposed two-stage adaptive enrichment design. Additionally, we seek to evaluate the efficacy of the proposed design. Three primary questions are addressed:
\begin{enumerate}
    \item Can the proposed methods accurately estimate the biomarker cutpoint in the proposed design?
    \item Are the proposed treatment effect estimators consistent with the true marginal RMST difference within the biomarker-positive subgroup in the proposed design?
    \item Does the proposed design effectively control the type I error rate while maintaining sufficient power? Additionally, what advantages does this design offer compared to the all-comer design?
\end{enumerate}

\subsection*{D.2: Data-Generating Mechanisms}
We simulate 10,000 datasets under two designs: 1) the proposed two-stage adaptive enrichment design and 2) the all-comer design with no enrichment.

The two-stage adaptive enrichment design follows the procedure outlined in Section 2.3. In Stage I, $n_1$ patients are accrued for each of the experimental and control treatments. The accrual in the first stage is regardless of the patient's biomarker value. We assume a uniform enrollment from calendar time 0 to $t_1 = 1$ year. For each patient $i$, the enrolled time $E_i \sim \text{Unif}(0,t_1)$. In Stage II, an additional $n_2$ patients are accrued per treatment. The accrual in Stage II is determined based on the enrichment strategy established at the end of Stage I. A uniform enrollment is assumed from $t_1$ to $t_2 = 2$ years. All patients will be followed up until $t_3 = 4$ years.

The all-comer design accrues and randomizes a total of $n=2\times(n_1+n_2)$ patients with one to one allocation regardless of their biomarker values. To follow a similar recruitment scheme as in the enrichment design, we assume that the first $2n_1$ patients are uniformly enrolled from time 0 to $t_1$ and the other $2n_2$ patients are uniformly enrolled from $t_1$ to $t_2$. All patients will be followed up until $t_3$.

In both designs, the event time $T_i$ for $Z \in \{0,1\}$ is generated from the following hazard functions: 
\begin{eqnarray}
    h_0(t|X) &=& 0.9, \nonumber \\
    h_1(t|X) &=& 
    \begin{cases}
      0.9 \times \text{exp}\{0.9(1-X)\} & \text{for $t \leq 0.25$} \\
      0.45 \times \text{exp}\{0.9(1-X)\} & \text{for $t > 0.25$},
    \end{cases}
\end{eqnarray}
Here, the time is measured in years. These hazard functions represent the scenario where the experimental treatment exhibits a delayed treatment effect after 3 months. We assume the lost to follow-up time $L_i$ follows an exponential distribution, specifically $L_i \sim \text{exp}(0.12)$. For an analysis conducted at time $t' \in \{t_1,t_3\}$, the censoring time for the $i^{th}$ patient is defined as $C_i = \text{min}(L_i, t' - E_i)$. eFigure 2 presents the true RMST curves by the biomarker values and their intersection point between the two treatments.

\begin{figure}[!htbp]
\centering\includegraphics[scale = 0.7]{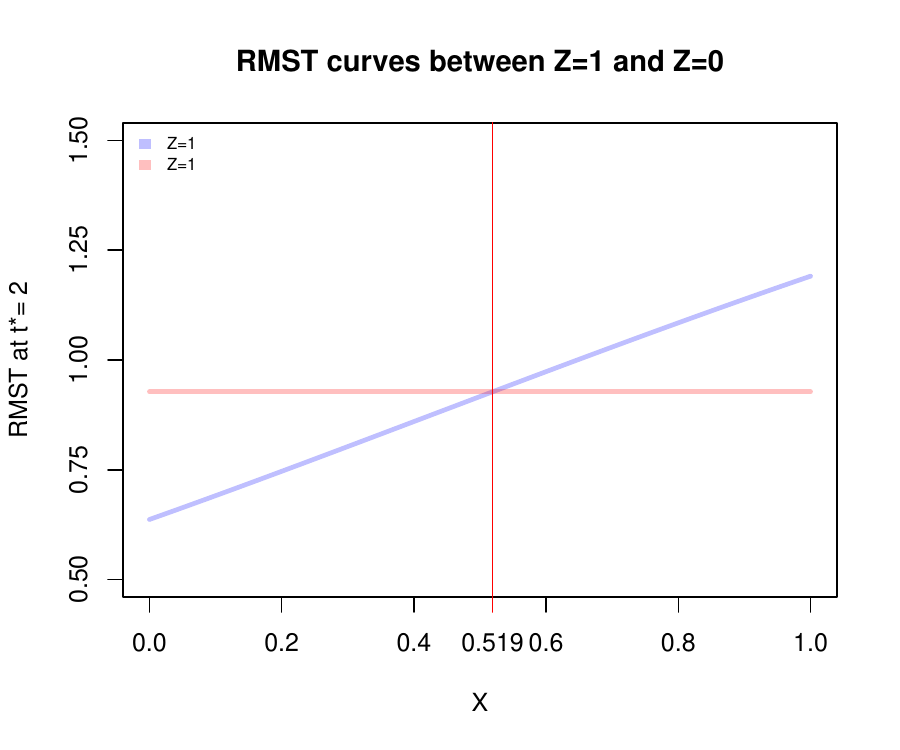}
\caption[RMST curves by biomarker in simulation study.]{RMST curves by the biomarker values for $Z=1$ and $Z=0$ in simulation study.}
\end{figure}

We use the RMST difference between experimental and control arms up to $t^*=2$ years to measure the treatment effect. Based on Equations (2.2) and (S.9), the true biomarker cutpoint is $c_{t^*} = 0.519$, and the true marginal RMST differences among the biomarker-positive subgroup $(c_{t^*}, 1]$ and overall patients are $\Delta_{t^*}^{(P)}=0.134$ and $\Delta_{t^*}^{(O)}=-0.012$, respectively.

We follow the same procedure as in the numerical example to calculate the sample size. Assuming equal accrual rates in Stages I and II, and the critical values $q=q_0=\Phi^{-1}(0.975)$, the total sample size required to achieve a global power of 80\% in an all-comer design using the naive unadjusted estimator is $n=2020$. As such, the sample size per stage and treatment arm is $n_1=n_2=505$.

\subsection*{D.3: Methods and Performance Measures}
The methods and performance measures to assess operating characteristics are summarized according to the three aforementioned primary questions and four design scenarios as described in Table 1.

\subsubsection*{Biomarker Cutpoint Estimation}
At the end of Stage I, we consider three different prediction methods to estimate the cutpoint according to the prior information on the piecewise exponential hazard functions:
\begin{description}
\item [Prediction Method A:] The number of change points in the piecewise exponential models and their locations are known.
\item [Prediction Method B:] The number of change points in the piecewise exponential models is known, but the locations are unknown.
\item [Prediction Method C:] Both the number of change points and their locations in the piecewise exponential models are unknown. The number of change points is tested under the significance level of 0.05.
\end{description}
We assess the bias of the estimated cutpoint, $\hat{c}_0$, in each prediction method and the accuracy of the detected number and locations of change points in the piecewise exponential models in Prediction Methods B and C.

At the end of Stage II, we use the RMST regression method to re-estimate the biomarker cutpoint. We assess the bias of the estimated cutpoint, $\hat{c}_{t^*}$, under all four design scenarios.

\subsubsection*{Treatment Effect Estimation}
We compare five treatment effect estimators, $\hat{\Delta}_l$ for $l=1,\dots,5$, for the marginal RMST difference in the biomarker-positive subgroup under four design scenarios. Each estimator is evaluated using both the estimated biomarker cutpoint $\hat{c}_{t^*}$ and the true fixed biomarker cutpoint $c_{t^*}=0.519$ to determine the biomarker-positive subgroup.

When estimating the marginal RMST difference using the estimated biomarker cutpoint $\hat{c}_{t^*}$, we assess the bias and coverage probabilities of each estimator. Coverage probability is defined as the proportion of replications in which 95\% confidence intervals contain the true value out of the total number of replications. The results are shown in the main manuscript.

When estimating the marginal RMST difference using the true biomarker cutpoint $c_{t^*}$, we evaluate the bias, coverage probability, estimated standard errors ($E[\hat{\sigma}_l]$) and the standard deviations ($\sqrt{Var[\hat{\Delta}_l]}$) of each estimator. This helps verify the correctness of the proposed variance of each estimator (i.e., $\hat{\sigma}_l^2$ for $l=1,\dots,5$). The results are shown in Web Appendix D.

\subsubsection*{Operating Characteristics}
We compare the operating characteristics in terms of global power, type I error rates and included true negative patients of the two-stage enrichment and all-comer designs under four design scenarios using five treatment effect estimators.

First, in the power analysis, we specify the significance levels for testing positive biomarker-by-treatment interaction and conditional treatment effect as $\alpha_0 = \Tilde{\alpha} = 2.5\%$, so that the associated critical values are $q=q_0=\Phi^{-1}(0.975)$. We compare the global power and number of included true negative patients in each design scenario and estimator, where the true negative patients are those whose biomarker values are within the range $[0, c_{t^*})$.

Second, the family-wise type I error rate is evaluated under the global null setting, and we also assess the type I error rate for testing positive biomarker-by-treatment interaction. We use the significance levels of $\alpha_0 \in \{1.5\%, 2.0\%, 2.5\%, 3.5\%\}$ for testing the positive biomarker-by-treatment interaction, and $\Tilde{\alpha} \in \{2.0\%, 2.3\%, 2.5\%\}$ for testing the conditional treatment effects. As a result, we present a total of 12 combinations of nominal significance levels. Under the global null setting, the Prediction Methods A and B are equivalent since there are no change points in both treatments. As such, we only present results of type I error rate under Design Scenarios 1, 2, and 4.

\subsection*{D.4: Additional Results}
eTable 1 presents the estimated RMST differences under four design scenarios using the fixed true biomarker cutpoint. Similar to the findings in Table 3, the Naive unadjusted estimator is unbiased in the all-comer design but is biased in the enrichment designs. In contrast, the four CW-adjusted estimators show unbiased results under all design scenarios. The estimated standard error of each estimator is close to its standard deviation under all design scenarios, with the associated coverage probabilities approximately 95\%. These results confirm the correctness of the variance estimators proposed in Web Appendix B.3. The CW G-formula estimator, $\hat{\Delta}_3$, demonstrates the smallest variance among the four CW-adjusted estimators.

eTable 2 presents the type I error rate for testing the positive biomarker-by-treatment interaction using four nominal significance levels of $\alpha_0$ under Design Scenarios 1, 2, and 4. As outlined in Section 5.1, under the global null, $P(Z_{\beta_3} > q_0) = \alpha_0$, where $q_0 = \Phi^{-1}(1-\alpha_0)$. Consequently, the type I error rate should align with the nominal significance level (i.e., $\alpha_0$). eTable 2 shows that type I error rates are effectively controlled in all design scenarios.

eTable 3 displays the family-wise type I error rates using 12 combinations of nominal significance levels of $\Tilde{\alpha}$ and $\alpha_0$ for each estimator under Design Scenarios 1, 2, and 4. Note that under the global null setting, the Prediction Methods A and B are equivalent since there are no change points in both treatments. First, under the same design scenario and treatment effect estimator, the family-wise type I error rate increases with higher $\alpha_0$ or $\Tilde{\alpha}$. Notably, the family-wise type I error rate is more sensitive to changes in $\Tilde{\alpha}$ than $\alpha_0$. Second, under the same significance levels and treatment effect estimator, the enrichment designs have slightly lower family-wise type I error rates than the all-comer design, and different prediction methods have minimal effect on the results. Finally, under the same design scenario and significance levels, the family-wise type I error rates are similar among the first three estimators. The CW Hajek and Augmented estimators ($\hat{\Delta}_4$ and $\hat{\Delta}_5$) provide lower family-wise type I errors than other estimators. However, this is attributed to slightly overestimating the variance in these two estimators under the global null setting (simulation results not shown). According to eTable 3, if we want to control the family-wise type I error under 0.025 in our proposed design using the CW G-formula estimator ($\hat{\Delta}_3$), one option is to set $\alpha_0=0.015$ and $\Tilde{\alpha}=0.023$.

\begin{table}[!htbp]
\caption{Estimated marginal RMST differences in the true biomarker-positive subgroup under four design scenarios, using the true biomarker-cutpoint $c_{t^*}$. RMSTD: marginal RMST difference in the true biomarker-positive subgroup, S.E. of RMSTD: estimated standard error of RMSTD, S.D. of RMSTD: standard deviation of estimated RMSTD's, C.P. of RMSTD: coverage probability of RMSTD. $\hat{\Delta}_1$: Naive unadjusted estimator; $\hat{\Delta}_2$: CW Kaplan-Meier estimator; $\hat{\Delta}_3$: CW G-formula estimator; $\hat{\Delta}_4$: CW Hajek estimator; $\hat{\Delta}_5$: CW Augmented estimator.}
\begin{center}
\begin{tabular}{ccccccc}
\hline
Design & Estimator & Est. & Bias of & S.E. of & S.D. of & C.P. of \\
Scenario & & RMSTD & RMSTD & RMSTD & RMSTD & RMSTD \\
\hline
1 & $\hat{\Delta}_1$ & 0.135 & 0.001 & 0.048 & 0.048 & 94.7\% \\
(All-Comer) & $\hat{\Delta}_2$ & 0.134 & -0.000 & 0.050 & 0.051 & 94.7\%\\
 & $\hat{\Delta}_3$ & 0.134 & -0.000 & 0.044 & 0.045 & 94.6\% \\
 & $\hat{\Delta}_4$ & 0.134 & -0.000 & 0.052 & 0.051 & 95.8\% \\
 & $\hat{\Delta}_5$ & 0.134 & -0.000 & 0.052 & 0.051 & 95.7\% \\
\hline
2 & $\hat{\Delta}_1$ & 0.144 & 0.010 & 0.040 & 0.040 & 94.5\% \\
(Enrichment) & $\hat{\Delta}_2$ & 0.133 & -0.001 & 0.044 & 0.046 & 94.8\%\\
 & $\hat{\Delta}_3$ & 0.133 & -0.001 & 0.038 & 0.039 & 94.7\% \\
 & $\hat{\Delta}_4$ & 0.133 & -0.001 & 0.046 & 0.046 & 95.7\% \\
 & $\hat{\Delta}_5$ & 0.133 & -0.001 & 0.046 & 0.046 & 95.6\% \\
\hline
3 & $\hat{\Delta}_1$ & 0.143 & 0.009 & 0.040 & 0.040 & 94.6\% \\
(Enrichment) & $\hat{\Delta}_2$ & 0.133 & -0.001 & 0.043 & 0.045 & 94.7\%\\
 & $\hat{\Delta}_3$ & 0.133 & -0.001 & 0.038 & 0.039 & 94.7\% \\
 & $\hat{\Delta}_4$ & 0.134 & -0.000 & 0.045 & 0.046 & 95.5\% \\
 & $\hat{\Delta}_5$ & 0.134 & -0.000 & 0.045 & 0.046 & 95.4\% \\
\hline
4 & $\hat{\Delta}_1$ & 0.145 & 0.011 & 0.041 & 0.042 & 93.5\% \\
(Enrichment) & $\hat{\Delta}_2$ & 0.134 & -0.000 & 0.045 & 0.048 & 94.6\%\\
 & $\hat{\Delta}_3$ & 0.134 & -0.000 & 0.038 & 0.040 & 94.4\% \\
 & $\hat{\Delta}_4$ & 0.134 & -0.000 & 0.047 & 0.048 & 95.3\% \\
 & $\hat{\Delta}_5$ & 0.134 & -0.000 & 0.047 & 0.048 & 95.2\% \\
\hline
\end{tabular}
\end{center}
\end{table}

\begin{table}[!htbp]
\caption[Type I error rate for testing $H_{00}$ in simulation study.]{Type I error rate for testing positive biomarker-by-treatment interaction. $\alpha_0$: significance level for testing Hypothesis Zero.}
\begin{center}
\begin{tabular}{cccc}
\hline
$\alpha_0$ & Design Scenario 1 & Design Scenario 2 & Design Scenario 4\\
\hline
0.015 & 0.016 & 0.016 & 0.016 \\
0.020 & 0.021 & 0.022 & 0.021 \\
0.025 & 0.026 & 0.027 & 0.027 \\
0.035 & 0.037 & 0.037 & 0.037 \\
\hline
\end{tabular}
\end{center}
\end{table}

\clearpage
\begin{table}[!htbp]
\caption[Family-wise type I error rate in simulation study.]{Family-wise type I error rate under each design scenario. $\alpha_0$: Significance level for testing positive biomarker-by-treatment interaction; $\Tilde{\alpha}$: Significance level for testing conditional treatment effect.}
\begin{center}
\begin{tabular}{cccccccc}
\hline
Design Scenario & $\alpha_0$ & $\Tilde{\alpha}$ & $\hat{\Delta}_1(t^*)$ & $\hat{\Delta}_2(t^*)$ & $\hat{\Delta}_3(t^*)$ & $\hat{\Delta}_4(t^*)$ & $\hat{\Delta}_5(t^*)$\\
\hline
1 & 0.015 & 0.025 & 0.029 & 0.029 & 0.028 & 0.024 & 0.024\\
(All-Comer) && 0.023 & 0.027 & 0.027 & 0.026 & 0.021 & 0.022\\
&& 0.020 & 0.024 & 0.024 & 0.022 & 0.019 & 0.019\\
&0.020 & 0.025 & 0.029 & 0.030 & 0.028 & 0.024 & 0.024\\
&& 0.023 & 0.027 & 0.028 & 0.026 & 0.022 & 0.022\\
&& 0.020 & 0.024 & 0.024 & 0.023 & 0.019 & 0.019\\
&0.025 & 0.025 & 0.030 & 0.031 & 0.030 & 0.025 & 0.025\\
&& 0.023 & 0.028 & 0.028 & 0.028 & 0.022 & 0.023\\
&& 0.020 & 0.025 & 0.025 & 0.024 & 0.019 & 0.019\\
&0.035 & 0.025 & 0.031 & 0.031 & 0.030 & 0.025 & 0.025\\
&& 0.023 & 0.029 & 0.029 & 0.028 & 0.023 & 0.023\\
&& 0.020 & 0.025 & 0.025 & 0.024 & 0.020 & 0.020\\
\hline
2 &0.015 & 0.025 & 0.029 & 0.029 & 0.027 & 0.023 & 0.023\\
(Enrichment) && 0.023 & 0.027 & 0.026 & 0.025 & 0.021 & 0.021\\
&& 0.020 & 0.024 & 0.023 & 0.022 & 0.018 & 0.018\\
&0.020 & 0.025 & 0.029 & 0.029 & 0.028 & 0.023 & 0.024\\
&& 0.023 & 0.027 & 0.027 & 0.026 & 0.021 & 0.022\\
&& 0.020 & 0.024 & 0.023 & 0.022 & 0.019 & 0.019\\
&0.025 & 0.025 & 0.030 & 0.030 & 0.029 & 0.024 & 0.024\\
&& 0.023 & 0.028 & 0.028 & 0.027 & 0.022 & 0.022\\
&& 0.020 & 0.025 & 0.024 & 0.023 & 0.019 & 0.019\\
&0.035 & 0.025 & 0.031 & 0.031 & 0.029 & 0.025 & 0.025\\
&& 0.023 & 0.029 & 0.028 & 0.027 & 0.022 & 0.023\\
&& 0.020 & 0.025 & 0.025 & 0.024 & 0.020 & 0.020\\
\hline
4 &0.015 & 0.025 & 0.029 & 0.028 & 0.027 & 0.023 & 0.023\\
(Enrichment) && 0.023 & 0.027 & 0.026 & 0.025 & 0.021 & 0.021\\
&& 0.020 & 0.023 & 0.023 & 0.022 & 0.018 & 0.018\\
&0.020 & 0.025 & 0.029 & 0.029 & 0.028 & 0.023 & 0.024\\
&& 0.023 & 0.027 & 0.027 & 0.026 & 0.021 & 0.022\\
&& 0.020 & 0.024 & 0.023 & 0.022 & 0.019 & 0.019\\
&0.025 & 0.025 & 0.030 & 0.030 & 0.029 & 0.024 & 0.024\\
&& 0.023 & 0.028 & 0.028 & 0.027 & 0.022 & 0.022\\
&& 0.020 & 0.025 & 0.024 & 0.023 & 0.019 & 0.019\\
&0.035 & 0.025 & 0.031 & 0.031 & 0.030 & 0.025 & 0.025\\
&& 0.023 & 0.028 & 0.028 & 0.028 & 0.022 & 0.023\\
&& 0.020 & 0.025 & 0.025 & 0.024 & 0.020 & 0.020\\
\hline
\end{tabular}
\end{center}
\end{table}